\documentclass{aa}  
\usepackage{natbib}
\bibpunct{(}{)}{;}{a}{}{,} 
\usepackage{graphicx}
\usepackage{array,tabulary,tabularx,multirow}
\usepackage{txfonts}
\usepackage{lipsum}
\usepackage[english]{babel}
\usepackage{color}
\usepackage{hhline}
\usepackage{multirow}
\usepackage{fourier}
\usepackage{booktabs}
\usepackage{threeparttable, tablefootnote}
\usepackage[dvipsnames]{xcolor}

\usepackage{romannum}
\begin{document}

   \title{ Searching for new variable white dwarfs: The discovery of the three new pulsating and three new binary systems}

   \subtitle{}

\author{Larissa Antunes Amaral
          \inst{1,2}, Maja Vučković\inst{1}, Ingrid Pelisoli\inst{3}, Alina Istrate\inst{4}, Kepler, S. O.\inst{5}, and Hibbert, Jacob M.\inst{6}
           }

   \institute{Instituto de Física y Astronomía, Universidad de Valparaíso, Gran Bretaña 1111, Playa Ancha, Valparaíso 2360102, Chile\\
   \email{larissa.amaral@postgrado.uv.cl}
              \and
              European Southern Observatory, Alonso de Cordova 3107, Santiago, Chile
              \and
             Department of Physics, University of Warwick, Gibbet Hill Road, Coventry, CV4 7AL, UK
             \and
             Department of Astrophysics/IMAPP, Radboud University, PO Box
            9010, 6500 GL Nijmegen, The Netherlands
            \and
             Instituto de Física, Universidade Federal do Rio Grande do Sul, 91501-970 Porto Alegre, RS, Brazil
             \and
             Isaac Newton Group of Telescopes (ING), Apto. 321, E-38700 Santa Cruz de la Palma, Canary Islands, Spain
             \\
               }

   \date{Received September 15, 1996; accepted March 16, 1997}

 
  \abstract{ 
  In recent years, approximately 150 low-mass white dwarfs (WDs), typically with masses below $0.4~M_{\odot}$, have been discovered. Observational evidence indicates that most of these low-mass WDs are found in binary systems, supporting binary evolution scenarios as the primary formation pathway. 
  A few extremely-low mass (ELM) WDs in this population have also been found to be pulsationally variable. In this work, we present a comprehensive analysis aimed at identifying new variable low-mass WDs. From our candidate selection we observed 16 objects which were identified within the ZZ Ceti instability strip. Those objects were observed over multiple nights using high-speed photometry from the SOAR/Goodman and SMARTS-1m telescopes. Our analysis led to the discovery of three new pulsating WDs: one pulsating ELM, one low-mass WD, and one ZZ Ceti star. Additionally, we identified three objects in binary systems, two with ellipsoidal variations in their light curves, one of which is likely a pre-ELM star, and a third showing reflection effect.

  } 

\titlerunning{New variables white dwarfs}
\authorrunning{L. Antunes Amaral et al.} 
   \maketitle
%

\section{Introduction}
White dwarf stars (WDs) are the final remnants of the evolution of all stars with an initial mass below $8.5$-$10 M_{\odot}$, and are thus the end-point of more than 95\% of the stars in the Milky Way \citep{Lauffer2018}. 
The observed sample of WDs covers a wide range of mass, ranging from as low as $\sim0.16 M_{\odot}$ to near the Chandrasekhar limit \citep[e.g.,][]{2016Tremblay,2017Kepler}. Specifically, for WDs with hydrogen (H) atmosphere (DAs), the mass distribution strongly peaks around $0.6M_{\odot}$ with fewer observed systems towards the low- and high-mass tails of the distribution. Those low- and high-mass WDs are most likely the result of close binary evolution, where mass transfer and mergers commonly occur. 


The extremely-low mass (ELM, $\lesssim $ 0.3$M_{\odot}$) and low-mass (LM, $\lesssim $ 0.45$M_{\odot}$) WD stars are expected to form only through binary interaction since such low-mass remnants (at least below $M<0.4M_{\odot}$) cannot be formed through single star evolution within Hubble time \citep{1995Marsh, zorotovic2017}.
The formation of these binary systems is believed to occur after an episode of enhanced mass loss in interacting binary systems, before helium is ignited at the tip of the red giant branch (RGB, \citep{2013Althaus, 2016Istrate, 2019Li}). 
This mass-loss can occur due to either (i) common-envelope evolution (CE) \citep{1976Paczynski, 1993IbenandLivio} or (ii) stable Roche-lobe overflow episode (RLOF) \citep{1986Iben}. In both cases, this interaction will leave behind an almost naked He remnant, which will later become a He WD \citep{2001Nelemans}, or a hybrid He-C/O WD if the mass is sufficient to ignite He after shedding the envelope. In the latter scenario, the star goes through a hot subdwarf phase before becoming a WD \citep{2019Zenati}.
Furthermore, short period binaries ($P < 6~h$) are formed through CE interaction, while binaries with longer orbital periods are likely formed through RLOF.

In close binary systems, reflection effects and ellipsoidal variations are common photometric phenomena. The reflection effect occurs when light from the primary star irradiates the companion’s surface, increasing the system's observed flux and producing quasi-sinusoidal light variations \citep{2022Barlow}. Ellipsoidal variations arise from tidal distortions, where one or both stars assume an ellipsoidal shape under the gravitational influence of the companion, leading to brightness modulations at half the orbital period \citep{2018Bell, 2022Barlow}.

The binary evolution scenario is currently supported by observations since most low-mass WDs are observed to be in binary systems \citep{2016Brown_ELM7}. The fraction of them that seems to be single is consistent with merger scenarios in close binary systems \citep{zorotovic2017}.

To date, about 150 ELM WDs are confirmed and characterised spectroscopically after a large progress was made in the study of those stars with the ELM Survey \citep[e.g.,][and references therein]{2022Brown_ELM9,2023ELMSouth2} and subsequent studies \citep{2019PelisoliandVos,Wang2022}. More than half of the known ELM WDs were discovered by analysing objects from the Sloan Digital Sky Survey (SDSS) \citep{2017Sdss_Blanton}, selected by colour cuts (see Figure 1 of \citet{2010Brown_ELM1}). Additionally, a few of these objects have been found to be variable ELMs (ELMV) through pulsations \citep{2012Hermes}. So far, eleven ELMVs have been observed in a low-mass extension of the hydrogen-atmosphere (DA) WD i.e. ZZ Ceti instability strip. The ELMVs were observed to pulsate with periods between 200~s and 7\,000~s \citep{ 2012Hermes, 2013Hermesa, 2013Hermesb, 2015Kilic, 2015Bell, 2017Bell,2018Pelisolib, 2021Guidry, 2021Lopez}, and to vary with low amplitudes of the order of milimagnitudes (mmag = ppt). As these amplitudes are typically too low for detection in large public photometric surveys such as TESS or ZTF, we conduct a dedicated follow-up to detect and characterize ELMV variability accurately.

The recent discovery of ELMVs has greatly sparked the interest in these objects, as it provides a unique opportunity to explore the internal structure of WDs at cooler temperatures (6.0 $\lesssim$ $\log g$ $\lesssim$ 6.8 and 7800~K  $\lesssim$ $T_{\mathrm{eff}}$ $\lesssim$ 10\,000~K) and much lower masses ($< $0.3$M_{\odot}$). The pulsating periods can probe the overall mass, rotation rate, convective efficiency, and, perhaps most importantly, the hydrogen envelope mass of these stars \citep{2008Winget_Kepler, 2008Fontaine_Brassard}. Therefore, the ELMVs have the potential to constrain the details of their interior structures and to better understand their formation histories through asteroseismology \citep{2010Aertsbook}.

Theoretical studies of the ELMV pulsations \citep{2010Steinfadt,2012Corsicod,a2014Corsico_Althaus,2018Calcaferro}, have shown that the gravity (\emph{g})-modes in ELMVs are mainly confined to the core regions, while the pressure (\emph{p}) and radial modes are confined to the stellar envelope regions.
It is believed that the $\kappa-\gamma$ mechanisms and convective driving, both acting in the H ionisation zone, excite the pulsations in these stars \citep{2012Corsicod,2013VanGrootel,2016Corscio_althaus}.
Additionally, it is also assumed that the $\varepsilon$ mechanism due to stable burning of H, could excite short-period \emph{g} modes in those stars \citep{a2014Corsico_Althaus}.

While asteroseismology offers the best probe of white dwarf interiors \citep{2008Winget_Kepler,2008Fontaine_Brassard}, the precise masses and radii measurements of ELMs can be obtained independently if they belong to an eclipsing binary system as well \citep{2017Parsons}. 

So far, there are four of known pulsating white dwarfs in detached eclipsing binary systems, one being a WD+MS system \citep{2015Pyzas}, the second being the first ELM WD in a double-degenerate system  \citep{2020Parsons}, and two latest discovery systems with pulsating WDs (ZZ Ceti) with companions of spectral type M and later (referred to as WD+dM) \citep{2023Brown}.  Given that WD in the mass range from 0.3 to 0.5$M_{\odot}$ are all expected to have a He-core, a low-mass carbon-oxygen or a hybrid core, those pulsating WD in eclipsing binaries are powerful benchmarks to constrain empirically the core composition of low-mass stellar remnants and investigate the effects of close binary evolution on the internal structure of WDs. 


Moreover, short-period binary white dwarfs are potential progenitors of thermonuclear supernovae, potentially including Type \Romannum{1}a when at least one of the components has a high enough mass \citep{1984Webbink,1984Iben_Tutukov,2007Bildsten}. Also, as strong sources of gravitational waves, ELMs will have an important contribution to the signal detected by space-based missions such as LISA \citep{2016Brown_ELM7,2017Amaro-Seoane_LISA}. 

In this work, we aim to expand the sample of known ELMV candidates by applying refined selection criteria based on Gaia DR2 data. Through targeted follow-up observations, we conduct a preliminary analysis of both pulsating and binary system stars to characterize these newly identified systems. This study sets the groundwork for future detailed analyses of their internal structures and evolutionary implications, contributing valuable benchmarks for understanding the formation and evolution of low-mass white dwarfs in binary systems.

\section{Candidate selection}

To search for pulsating ELM WDs, we commence with the Gaia Data Release 2 Catalogue of Extremely-low Mass White Dwarf Candidates \citep{2019PelisoliandVos}. Within this catalogue, the authors initially mapped the parameter space occupied by known ELMs in the Gaia observational HR diagram, correlating absolute magnitude $G_{abs}$ with the colour $G_{BP}$ -- $G_{RP}$. The resulting catalogue of known ELMs is depicted in Table 1 of \citet{2019PelisoliandVos}, comprising 119 objects. Most of the catalogued ELMs are situated between the main sequence and canonical mass white dwarfs, consistent with their intermediate radii.

Subsequently, the authors analysed the position of the known ELMs with masses M $<$ 0.3 $M_{\odot}$ and parallax\_over\_error $>$ 5, along with the position of an evolutionary model corresponding to the upper mass limit as shown in Figure 1 of \citet{2019PelisoliandVos}.  Establishing a set of colour cuts and quality flags (see \citet{2019PelisoliandVos} for further details), the authors identified 5762 ELM candidates.

To assemble a complete, all-sky sample of ELM candidates with measured distances, we initiated a selection process based on known ELMs’ parameter space and theoretical considerations on the Gaia HR diagram ($M_G$ versus $G_{BP}$ -- $G_{RP}$). This selection adheres to recommended Gaia DR2 quality criteria \citep{2018Lindegrenetal, 2019Boubert} and includes colour cuts to reduce contamination by canonical white dwarfs and cataclysmic variables. Efforts are ongoing to characterise these candidates via multi-epoch spectroscopy, with the ultimate goal of compiling a volume-limited sample of ELMs that could provide a benchmark for binary evolution models. About 300 stars within 350 pc have been observed so far.

\begin{figure}[h!]
   \centering
   \includegraphics[width=0.45\textwidth]{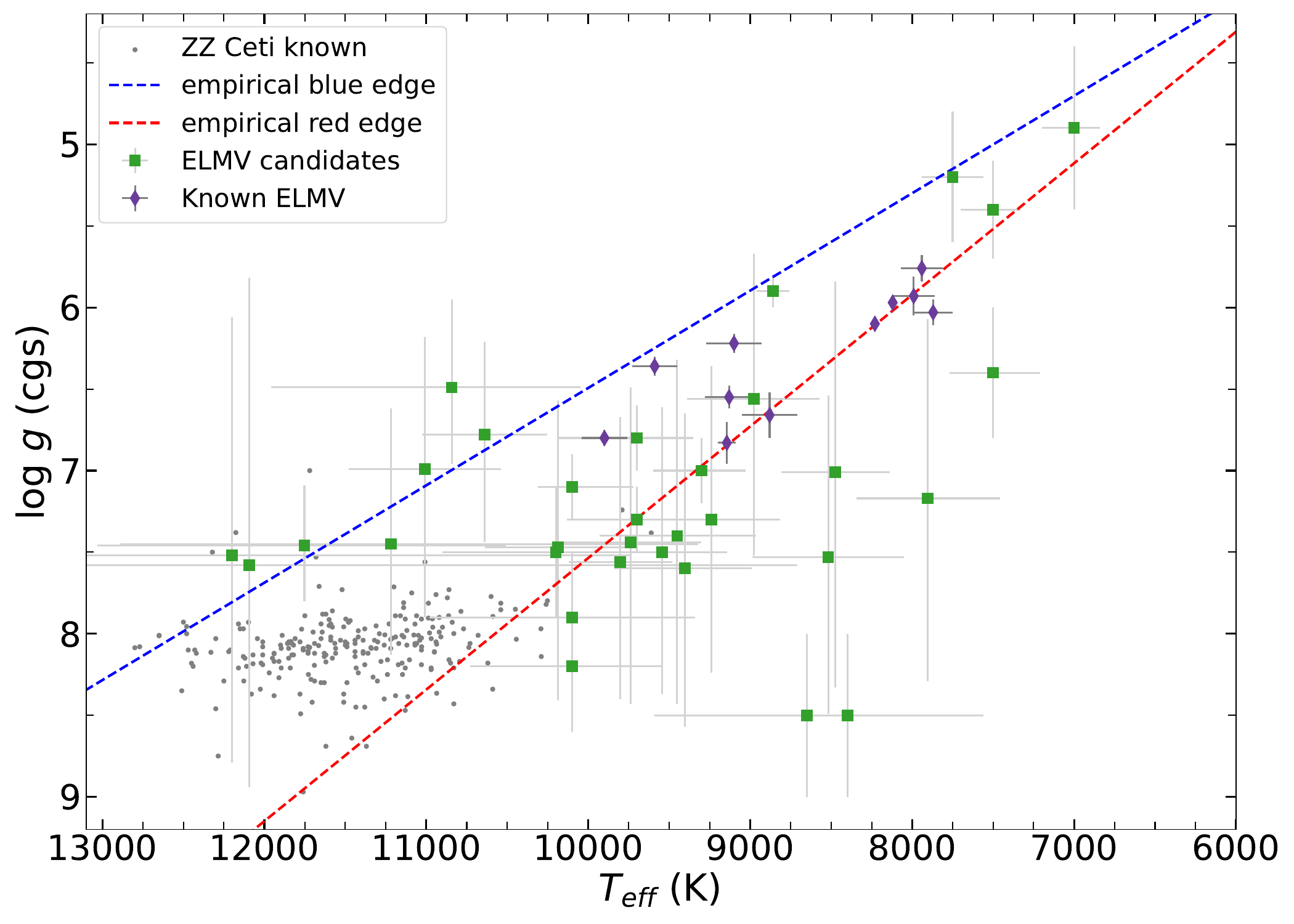}
      \caption{The position of the candidates sample are shown as green squares in the $T_{\mathrm{eff}}-\log{g}$ plane. The known ZZ Ceti \citep{2022MRomero} and ELMVs \citep{2012Hermes, 2013Hermesa, 2013Hermesb, 2015Kilic, 2015Bell, 2017Bell, 2018Pelisolib, 2021Lopez} are shown as grey dots and purple diamond shapes, respectively. The ELMV found by \citet{2021Guidry} is not being depicted since its atmospheric parameters have not been determined. The empirical ZZ Ceti instability strip published in \citet{2015Gianninas_ELM6} is marked with dashed lines.
      }
\label{Fig:Teff_Logg_ELMVcand}
\end{figure}

Leveraging this ongoing catalogue, we have utilised stars from this sample with existing spectral fits, relying on their preliminary $T_{\mathrm{eff}}$ and $\log g$ values, plotting them in a $T_{\mathrm{eff}}$ \textit{versus} $\log g$ diagram to assess their location. Subsequently, we selected those within the instability strip, accounting for their error bars, regardless of their mass. Consequently, our sample can include canonical WDs, low-mass WDs, and ELM WDs. We observed a total of 16 pulsating candidates (green squares in Figure \ref{Fig:Teff_Logg_ELMVcand}).


\section{Observations}

\subsection{High speed photometry}
High-speed time-series photometry was obtained using  Goodman spectrograph at the 4.1~m Southern Astrophysical Research (SOAR) Telescope and The Small and Moderate Aperture Research Telescope System (SMARTS) 1 meter telescope. With Goodman/SOAR we used the Blue Camera with the S8612 red-blocking filter. We used read-out mode 200 Hz ATTN2 with the CCD binned 2$\times$2 and integration time ranging from 4 to 80 seconds, depending on each star's magnitude and weather conditions, and with about 6\,s dead time between each exposure. For SMARTS telescope we used Apogee F42 camera and the SDSSg filter, with the CCD binned 1$\times$1 and integration time from 28 to 60 seconds, with about 4\,s dead time. Further photometric observations details can be seen in Table \ref{table:phot_log}.

Both SOAR and SMARTS photometric data were reduced using the software {\tt IRAF}, with the package \textsc{DAOPHOT} to perform aperture photometry. All photometry images were bias-subtracted, and flat field corrected using dome flats. Neighbouring non-variable stars of similar brightness were used as comparison stars to perform the differential photometry. We, then, divided the light curve of the target star by the mean light curve of all comparison stars to minimise the effects of sky and transparency fluctuations. 


\subsection{Spectroscopy}
\label{SUBSection_spectroscopy}

The ongoing characterisation of ELM candidates is being carried out with multiple telescope and instruments to ensure all-sky coverage. Five out of the 16 stars in this work have SDSS spectra (from the original SDSS spectrograph). Six were observed with the Goodman spectrograph \citep{2004Clemens_SOAR} at SOAR, four with the Intermediate Dispersion Spectrograph (IDS) installed at the 2.54 m Isaac Newton Telescope (INT), and the remaining one was observed with the Gemini Multi-Object Spectrographs (GMOS) at the 8-m Gemini South telescope \citep{Hook2004, GMOS-South}. Details about the configuration at each telescope are shown in Table \ref{table:spec}. For all observations, we aimed at covering the optical range, in particular the region of high-order Balmer lines, which is sensitive to $\log~g$ in the low-mass range, and the slit width was set to be comparable to the seeing. 

All the data were reduced using {\tt IRAF}. We performed bias subtraction, and flat-field correction using internal flat lamps. Arc lamp spectra for wavelength calibration were taken for every observation with the same pointing to minimise flexure effects. Flux calibration was performed with a standard star observed with the same setup. Due to slit losses the flux calibration is not absolute, but corrects for the instrument response.





\section{Methods}

\subsection{Light curve analysis: looking for periodicities}
\label{subsec:lc_analysis}
\begin{figure}[ht!]
\begin{center}
  \includegraphics[width=0.45\textwidth]{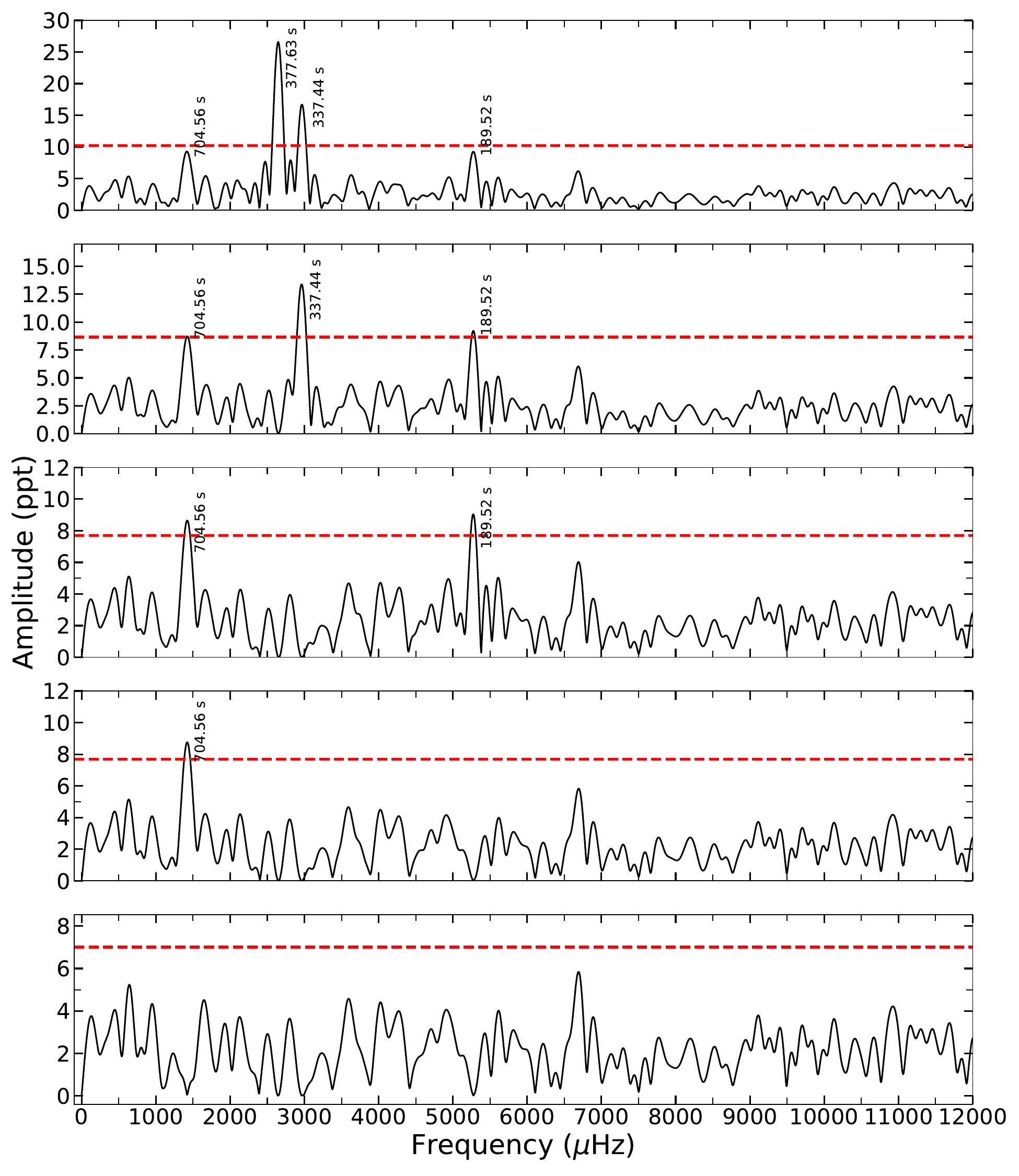}
  \caption{Pre-whitening process for the star SDSS J090559.60+084324.9. All real peaks above the detection threshold (depicted as the red dashed line) were subtracted until only the noise from the FT remained.}
    \label{Fig:pre-whitening}
  
\end{center}
\end{figure}

To look for periodicities in the light curves due to pulsations, we calculate the Fourier Transform (FT), using the software Period04 \citep{2004Lenz_Breger}.  We accepted a frequency peak as significant if its amplitude exceeds an adopted significance threshold. In this work, the detection limit (e.g. dashed line in Figure \ref{Fig:pre-whitening}) corresponds to the 1/1000 False Alarm Probability (FAP), where any peak with amplitude above this value has 0.1\% probability of being a false detection due to noise. The FAP is calculated by shuffling the fluxes in the light curve while keeping the same time sampling, and computing the FT of the randomised data. This procedure is repeated N/2 times, where N is the number of points in the light curve. For each run, we compute the maximum amplitude of the FT. From the distribution of maxima, we take the $0.999$ percentile as the detection limit \citep{2020Romero_hotdav}. Also, for each new FT and LC after each pre-whitening process, we calculate a new value for the FAP. The internal uncertainties in frequency and amplitude were computed using a Monte Carlo method with 100 simulations with \texttt{PERIOD04}, while uncertainties in the periods were obtained through error propagation.

\subsection{Spectroscopic fit}

\begin{figure}[ht!]
\begin{center}
  \includegraphics[width=\columnwidth, angle=0]{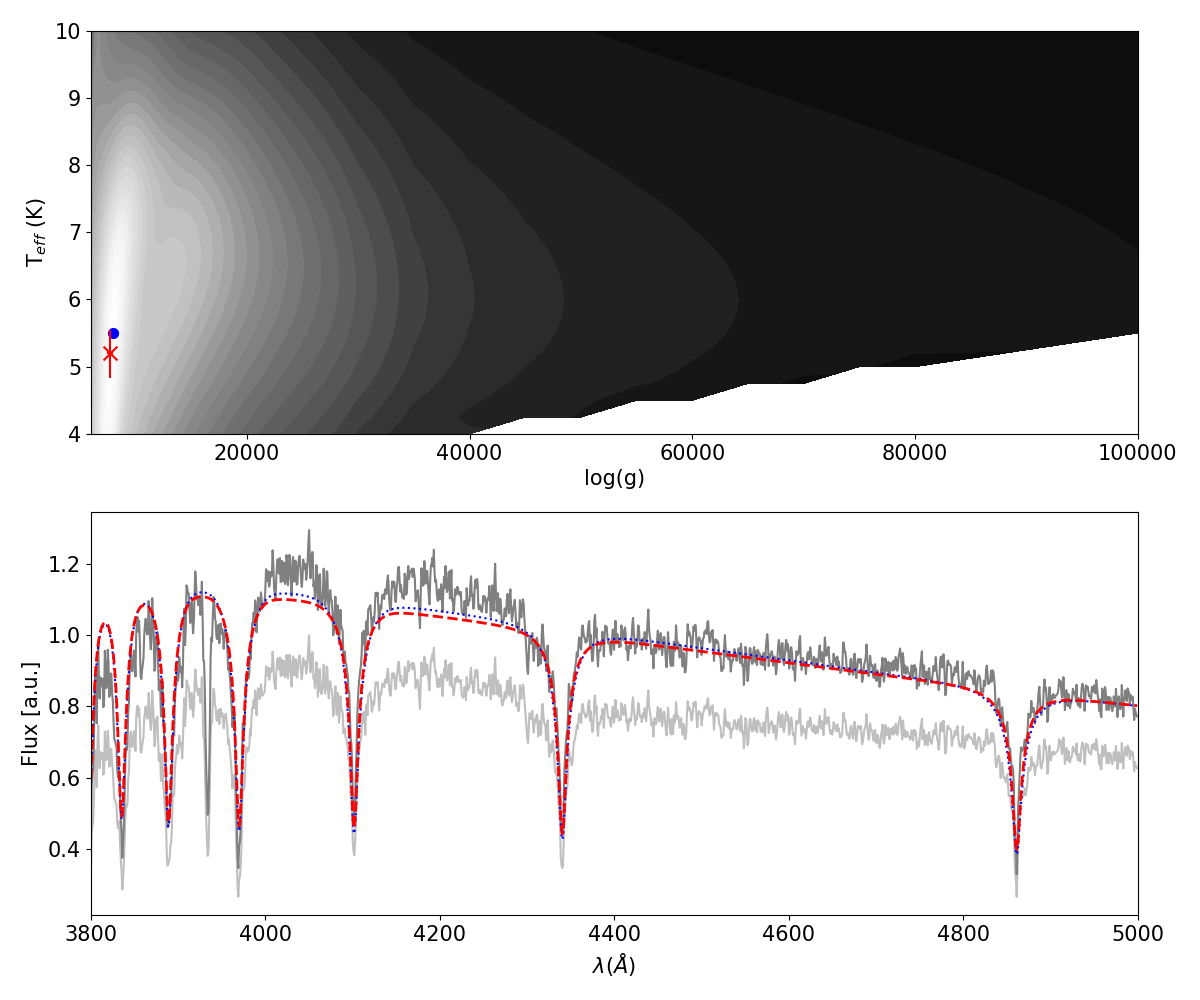}
  \caption{Spectroscopic fit of J183702.03-674141.1. The top panel shows the $\chi^2$ as a function of $T_{\mathrm{eff}}$  and $\log~g$, with the blue dot indicating the absolute minimum. The red error bar shows the median (marked by a cross) and 68\% confidence interval. The bottom panel shows the observed spectrum in light grey, the extinction-corrected spectrum in darker grey, the model for minimum $\chi^2$ in blue, and the adopted solution in red. Note that this object has atmospheric metals which are not included in the model.}
    \label{Fig:chi2fit}
\end{center}
\end{figure}

\begin{figure}[ht!]
\begin{center}
  \includegraphics[width=\columnwidth, angle=0]{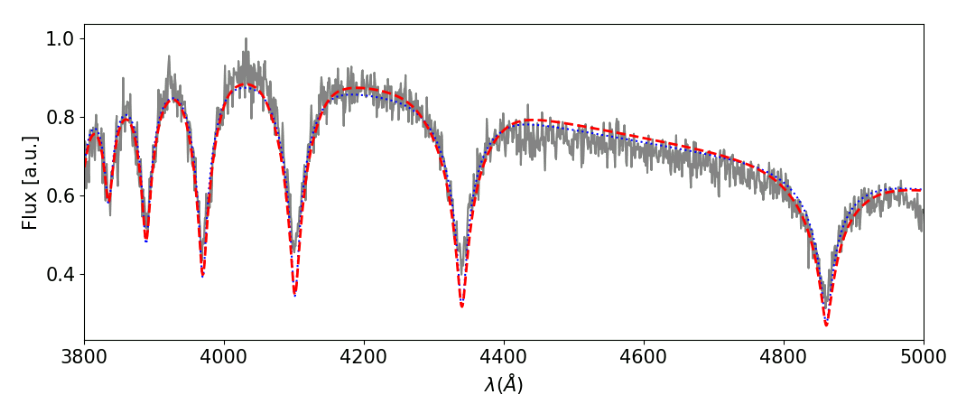}
  \includegraphics[width=0.7\columnwidth, angle=0]{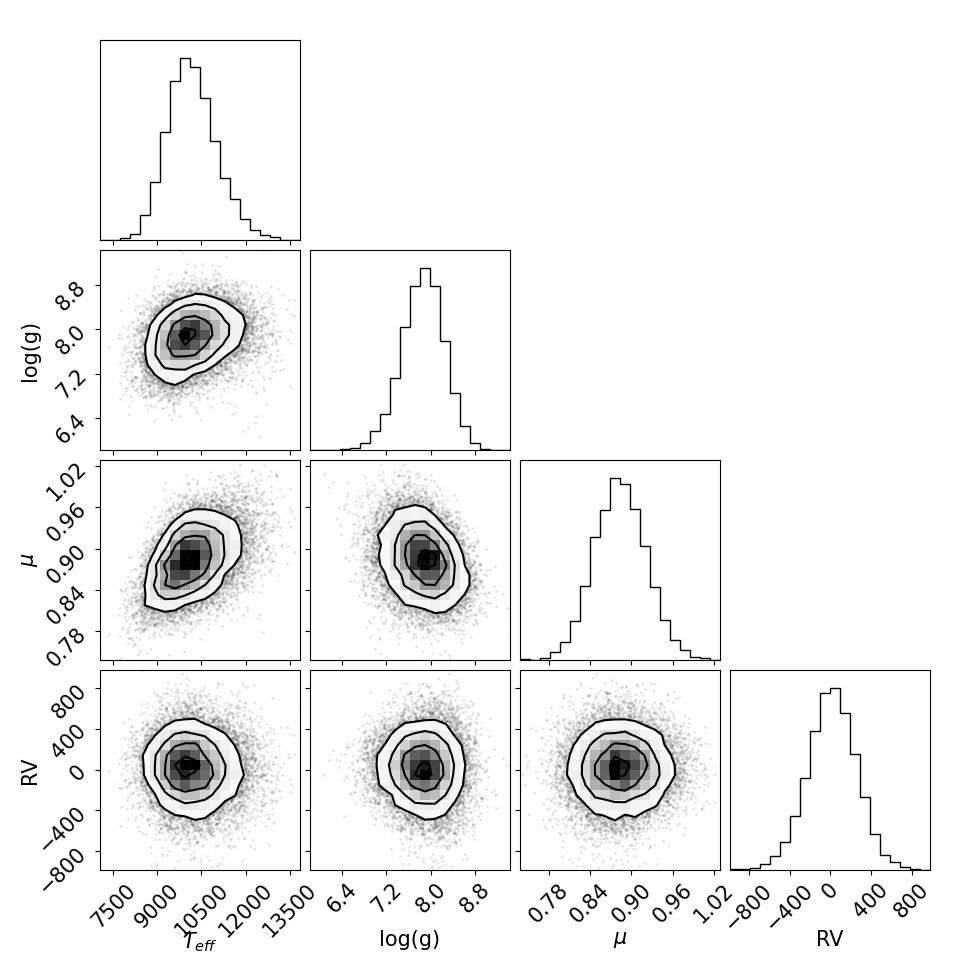}
  \caption{Spectroscopic fit of J151626.39-265836.9. The top panel shows the observed spectrum in light grey, the extinction-corrected spectrum in darker grey (with large overlap due to low extinction), the model for minimum $\chi^2$ in blue, and the adopted solution in red. The bottom panel shows the corner plot for the MCMC fit, where RV is the radial velocity in km/s and $\mu$ is the constantant normalisation parameter.}
    \label{Fig:mcmcfit}
\end{center}
\end{figure}

The spectra were fitted in the range 3800--5000~\AA\ using a pure-hydrogen atmosphere grid similar to that used in \citet{Kepler2021}, with the $\log~g$ range extended down to 4.0. For each observed spectrum, we calculated a $\chi^2$ with every model in the grid, first broadening the models to the resolution of the observations, and applying an extinction-correction to the observed spectrum using the $A_V$ values of \citet{Lallement2019Stilism} at each target's {\it Gaia}-derived distance and \citet{Fitzpatrick2007}'s extinction law. The spectra and models were normalised by a constant free parameter, preserving the slope of the observations but not the absolute flux, given potential slit losses. The radial velocity was also left as a free parameter. This resulted in an absolute minimum for the majority of targets (see example in Fig.~\ref{Fig:chi2fit}), with a few exceptions where there were two possible solutions. In these cases, we ran a Markov-Chain Monte Carlo (MCMC) fit with the same free parameters, but placing a prior on the effective temperature to make it consistent with the temperature derived from a spectral energy distribution (SED) fit to the SDSS and {\it Gaia} magnitudes (see example fit in Fig.~\ref{Fig:mcmcfit}). Uncertainties were determined from the 68\% confidence interval of the $\chi^2$-determined likelihood for the $\chi^2$ fits, or from the posterior distribution in the MCMC fits. The results of the fits are reported in Table~\ref{table:variabilityInfo}, and the fits not shown in the main text are in Appendix~\ref{appendix:spectra}, Figs. \ref{Fig:fit_mcmc} (MCMC fit) and \ref{Fig:fit_chi} ($\chi^2$ fit). In some cases, the reduced $\chi^2$ of the fit is high ($\gtrsim 50$); that is a consequence of one of two scenarios: i) the presence of metals, which remain in the atmosphere of low-mass WDs for longer due to rotational mixing \citep{2016Istrate}, or ii) the presence of an unseen companion contributing significantly to the flux. When metals are present, there is an additional systematic uncertainty of up to 1~dex in $\log~g$ \citep[see][]{2018Pelisolia}. The presence of a companion is inferred by the occurrence of extra light in the spectrum, which can be noticed from a spectral slope that cannot be described by the models (J0159-1805, J0406-5427, J1832+1413) or from a mismatch between the spectral slope and width of the spectral lines (J1346-1350, J2129). The former is more likely caused by a companion with a different colour, whereas the latter by a companion with a similar colour that only dilutes the absorption lines. Determining the exact nature of the companion would require either larger spectral coverage, as spectra for the systems showing indication of a companion do not extend beyond $\sim 5000$~\AA, or radial velocity monitoring. As neither metals or the dilution by a companion are taken into account here, the reported values in Table~\ref{table:variabilityInfo} should be interpreted with caution, but serve their purpose of identifying potential systems in the instability strip for follow-up. 

\section{Results}
We report the discovery of three new pulsating white dwarfs: one potential pulsating Extremely Low-Mass (ELMV) WD, one low-mass white dwarf, and one ZZ Ceti star. All of which are discussed with further details in Section \ref{Section_new_pulsating}).
Additionally, in our search for variability, we identified three stars in binary systems. Among these, two exhibit ellipsoidal variations; one of these is likely a pre-ELM star, while the third shows reflection effects (see Section \ref{Section_new_binary}). Detailed information on the effective temperatures, surface gravity, masses, and types of variations for these targets is provided in Table \ref{table:variabilityInfo}.

Furthermore, we observed no detectable variability in 10 other stars in our sample (see Section \ref{Section_nov}).

\begin{figure}[ht!]
\begin{center}
  \includegraphics[width=0.9\columnwidth, angle=0]{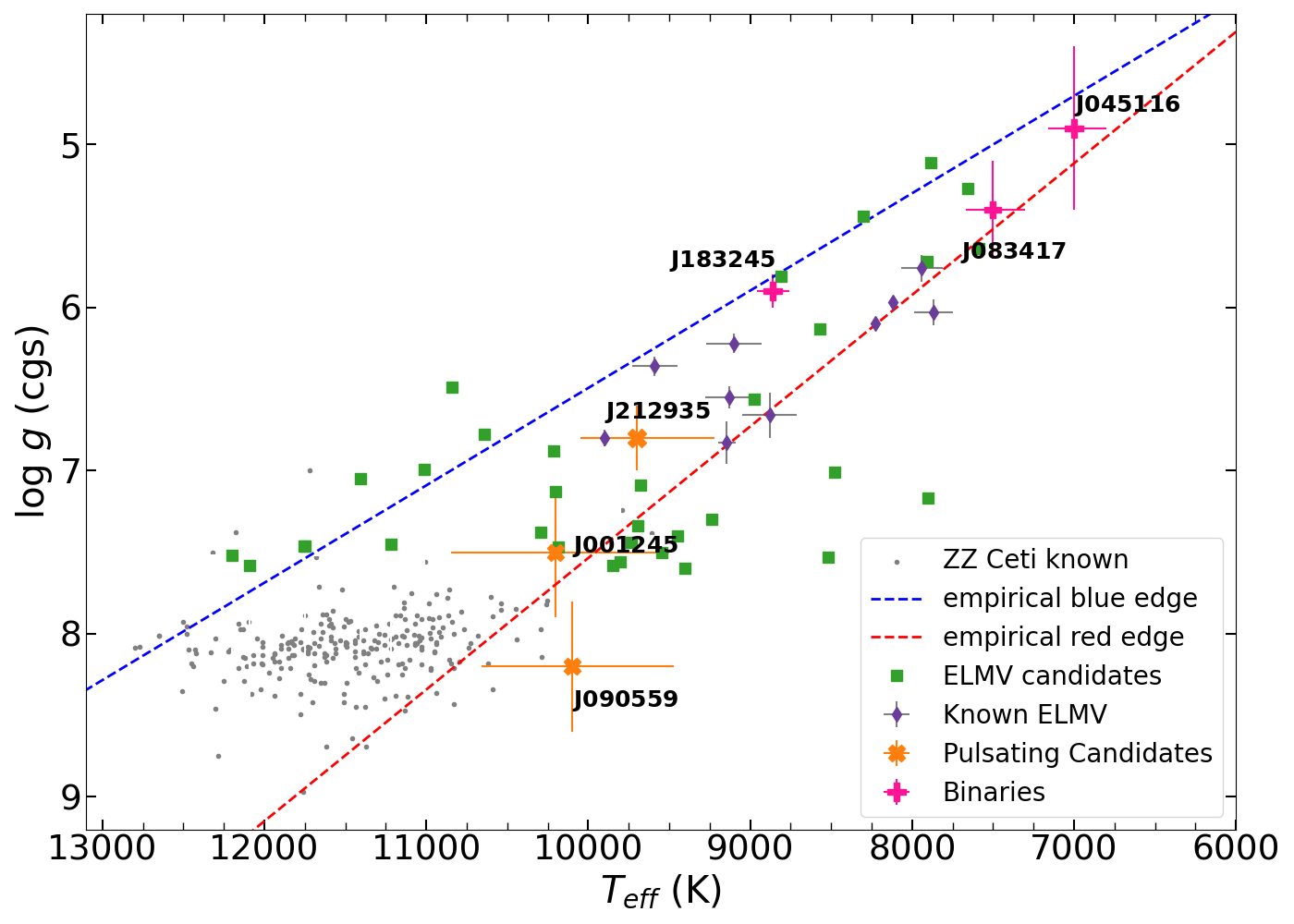}
  \caption{$T_{\mathrm{eff}}-\log{g}$ plane depicting the position of the three new pulsating WDs, shown as orange $\times$ symbols, and the three new binary systems, shown as pink $+$ sign. All values for $T_{\mathrm{eff}}$ and $\log{g}$ for these six new variable objects are those calculated in this work. The known ZZ Ceti \citep{2022MRomero} and ELMVs \citep{2012Hermes, 2013Hermesa, 2013Hermesb, 2015Kilic, 2015Bell, 2017Bell, 2018Pelisolib, 2021Lopez} are shown as black dots and purple diamond shapes, respectively. The empirical ZZ Ceti instability strip published in \citet{2015Gianninas_ELM6} is marked with dashed lines.}
    \label{Fig:TeffLogg_with_stars_marked}
  
\end{center}
\end{figure}

\renewcommand{\arraystretch}{1.7}

\subsection{New pulsating white dwarfs}
\label{Section_new_pulsating}

\subsubsection{SDSS J212935.23+001332.3} 

SDSS J212935.23+001332.3, with a magnitude of $G=15.54$ and located at a Gaia-derived distance of 1/parallax=$65.4 \pm 0.2$ pc, was observed on three separate nights with the SOAR telescope and once with the SMARTS 1m telescope, resulting in a total of approximately 9.5 hours of observations. Due to intermittent cloudy conditions on some nights, not all data could be used. To ensure the accuracy of our final analysis, only observations from nights with clear conditions were included for a thorough FT analysis. 

Our spectral fitting of the observed INT spectra yields an $T_{\mathrm{eff}}$ of 9700$^{+480}_{-350}$ K , $\log g$ = 6.8$\pm$0.2 consistent within 2.4-$\sigma$ with the values reported by \citet{2023Caron} of $T_{\mathrm{eff}}$=8860 $\pm$ 30 K, $\log g$=7.16 $\pm$ 0.006, along with a measured mass of 0.244 $\pm$ 0.002 \(\mathrm{M}_{\odot}\) for the same star (see Figure \ref{Fig:logg_teff-pulsation_comp_append}). The spectrum also shows indications of potential excess light that could originate from a binary companion. However, further spectral observations and RV measurements are required to confirm the binary nature of this object.  

\begin{figure}[ht!]
\begin{center}
  \includegraphics[width=8cm, angle=0]{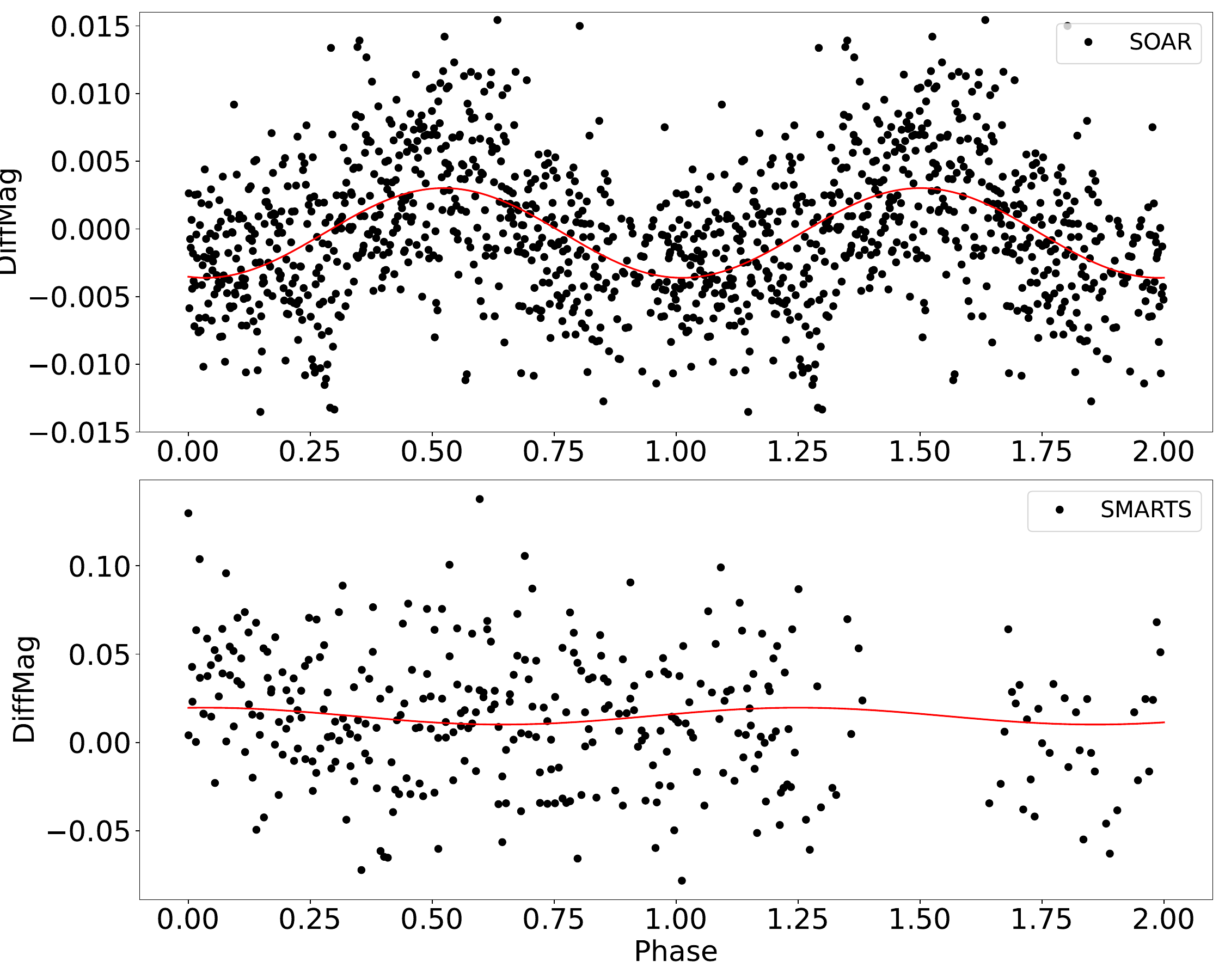}
  \caption{ Phased-folded light curve for the new ELMV SDSS J212935.23+001332.3 for the first observing night with SOAR telescope (upper panel) and SMARTS--1m telescope (bottom panel). Both light curves were folded to the highest amplitude pulsational period of about P=$1.22$ hours. The red line indicates a sinusoidal fit, depicted for better visualisation of the variation. }
    \label{Fig:Gaia2687_lc_smarts_soar}
  
\end{center}
\end{figure}

Figure \ref{Fig:Gaia2687_lc_smarts_soar} displays the phase-folded light curves from one of the SOAR nights (top panel) and the SMARTS night (bottom panel), with sinusoidal fits indicated by the red lines. Due to unstable weather conditions, smaller SNR, and the longer cadence of the SMARTS observations, the SMARTS light curve does not clearly exhibit the sinusoidal pattern visible in the SOAR light curve, as shown in the same figure.

\begin{figure}[ht!]
\begin{center}
  \includegraphics[width=8cm, angle=0]{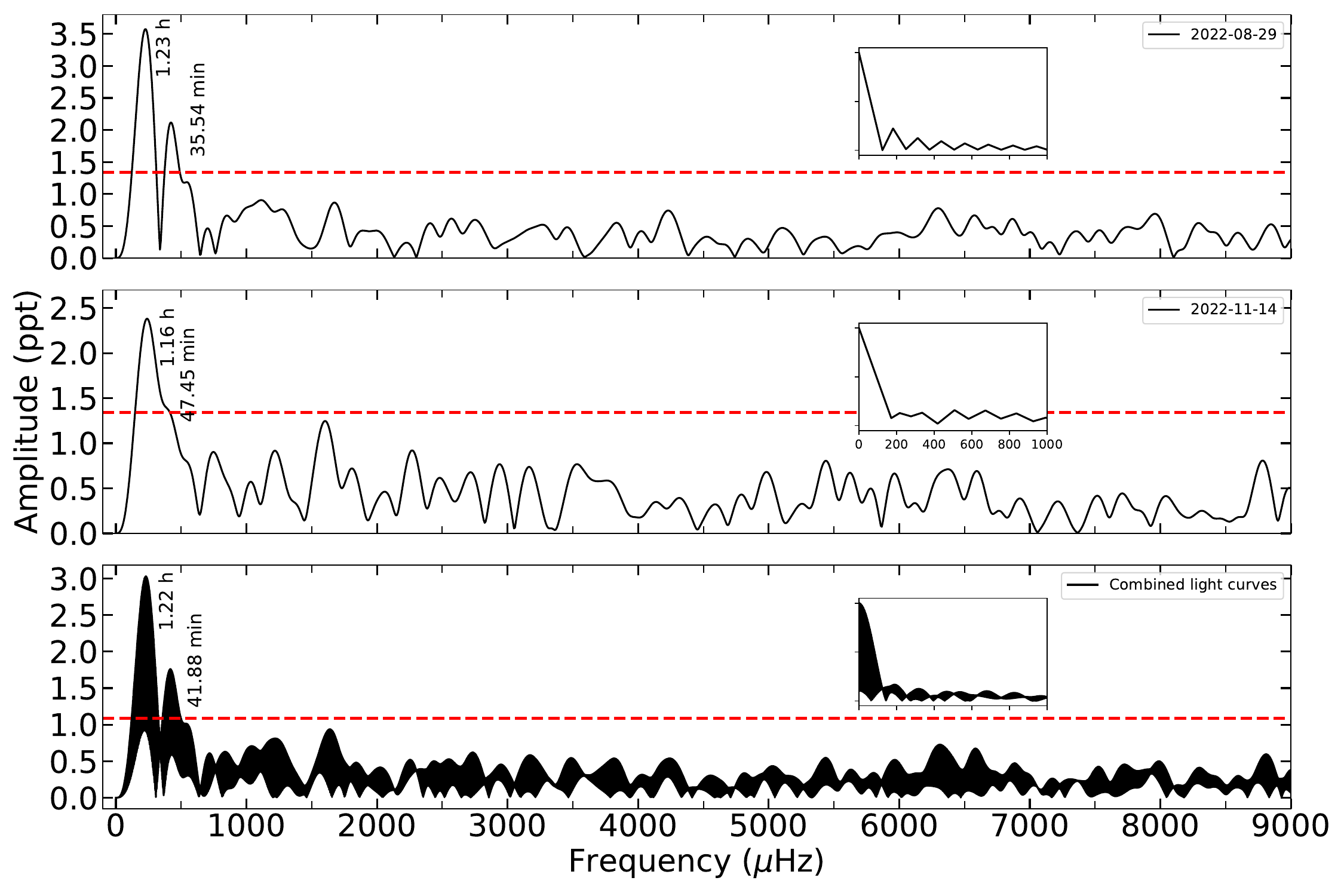}
  \caption{FT for the new ELMV SDSS J212935.23+001332.3. The second and fourth observed SOAR nights are shown in the top and middle panels, respectively, while the combined data from those both nights are displayed in the bottom panel. The amplitude at FAP = 1/1000 detection limit in each case is indicated by the horizontal red dashed line, and the spectral window for each case is depicted as an inset plot. }
    \label{Fig:newELMVFT}
  
\end{center}
\end{figure}

The FT of the SOAR light curves is presented in Figure \ref{Fig:newELMVFT}. In this figure, the data from the second and fourth nights are shown in the top and middle panels, respectively, while the bottom panel displays the combined data from these two nights. Unfortunately, the unstable cloud conditions during the first and third SOAR nights, as well during the night of observations with SMARTS, prevented a reliable analysis of the pulsation peaks in the FT, and thus their data are not included here. The horizontal red dashed line in all panels of Figure \ref{Fig:newELMVFT} indicates our threshold for FAP of 1/1000.

Our analysis identified two prominent peaks above FAP=1/1000 threshold in the FT, corresponding to periods of approximately 1.22 hours and 42 minutes. These periods are highly consistent with the typical pulsation periods of ELMVs, which range from 200~s to 7\,000~s \citep{2016Corscio_althaus, 2016Istrate, 2018Calcaferro}. Details about these peaks are provided in Table \ref{table:ELMV_new_freq_period_amp}. Furthermore, this star is placed close to the known ELMVs (see Figure \ref{Fig:TeffLogg_with_stars_marked}) in the $T_{\mathrm{eff}}$--$\log g$ plane, combined with the calculated mass 0.244  \(\mathrm{M}_{\odot}\) \citep{2023Caron}. The combination of the pulsation periods, observed in this object, characteristic of ELMVs, and the $T_{\mathrm{eff}}$ and $\log g$ values determined in this work, suggests that this object is a compelling new addition to the known population of pulsating ELM WDs.


\begin{table*}[h!]
\caption{Detected frequencies for the new ELMV SDSS J212935.23+001332.3.}             
\label{table:ELMV_new_freq_period_amp}      
\centering          
\begin{tabular}{c c c c c }     
Observing date & Freq ($\mu Hz$)&	Amp (ppt)	&	P (hr)	&	FAP	(ppt)\\ 
 \hline\hline
\multirow{2}{*}{2022-08-29}   &225.079$\pm$0.6 &3.6$\pm$0.3$\pm$ & 1.234$\pm$0.003 & 1.34 \\
&468.915$\pm$11& 1.6$\pm$0.3$\pm$ & 0.59$\pm$0.01 & 1.010 \\ \hline
\multirow{2}{*}{2022-11-14}   &239.133$\pm$15&2.4$\pm$0.4  & 1.16$\pm$0.07 & 1.34 \\
&351.227$\pm$18 &  1.5$\pm$0.5  & 0.79$\pm$0.04 & 1.28 \\ \hline
\multirow{2}{*}{Nights combined}   &227.042$\pm$0.004 &3.0$\pm$0.2  & 1.21904$\pm$0.00002 & 1.08 \\
& 397.959$\pm$0.009& 1.3$\pm$0.2  & 0.69783$\pm$0.00002 & 1.025 \\ \hline
\end{tabular}

{\raggedright \textbf{Notes:} The table lists the observing night, frequency, amplitude, period, and FAP detection limit for each peak. Observations were performed with SOAR. \par}
\end{table*}


As shown in Figure \ref{Fig:TeffLogg_with_stars_marked}, SDSS J212935.23+001332.3 is located within the instability strip as well as close to the known ELMVs.

\subsubsection{SDSS J001245.60+143956.4} 

SDSS J001245.60+143956.4 was observed during 3.4hrs with SOAR. With a magnitude of $G=18.19$ and a Gaia-derived distance of 1/parallax=$359 \pm 24$ pc, this object has also an SDSS spectrum available. Our spectral fitting gave us a $T_{\mathrm{eff}}$=10200$^{+700}_{-650}$K and $\log g$=7.5$\pm 0.4$, being perfectly consistent with the existing solution from \citet{2013Kleinman} , in which the authors have estimated $T_{\mathrm{eff}}$=10893 $\pm$ 68K, $\log g$=7.74 $\pm$ 0.07 and a mass of 0.482 $\pm$ 0.03 ${M}_{\odot}$ (see Figure \ref{Fig:logg_teff-pulsation_comp_append}). For this object the measured $T_{\mathrm{eff}}$ and $\log g$ agree with each other within uncertainties, and \citet{2013Kleinman} delivered mass places this objects as a low-mass WD. 

\begin{figure}[ht!]
\begin{center}
  \includegraphics[width=8cm, angle=0]{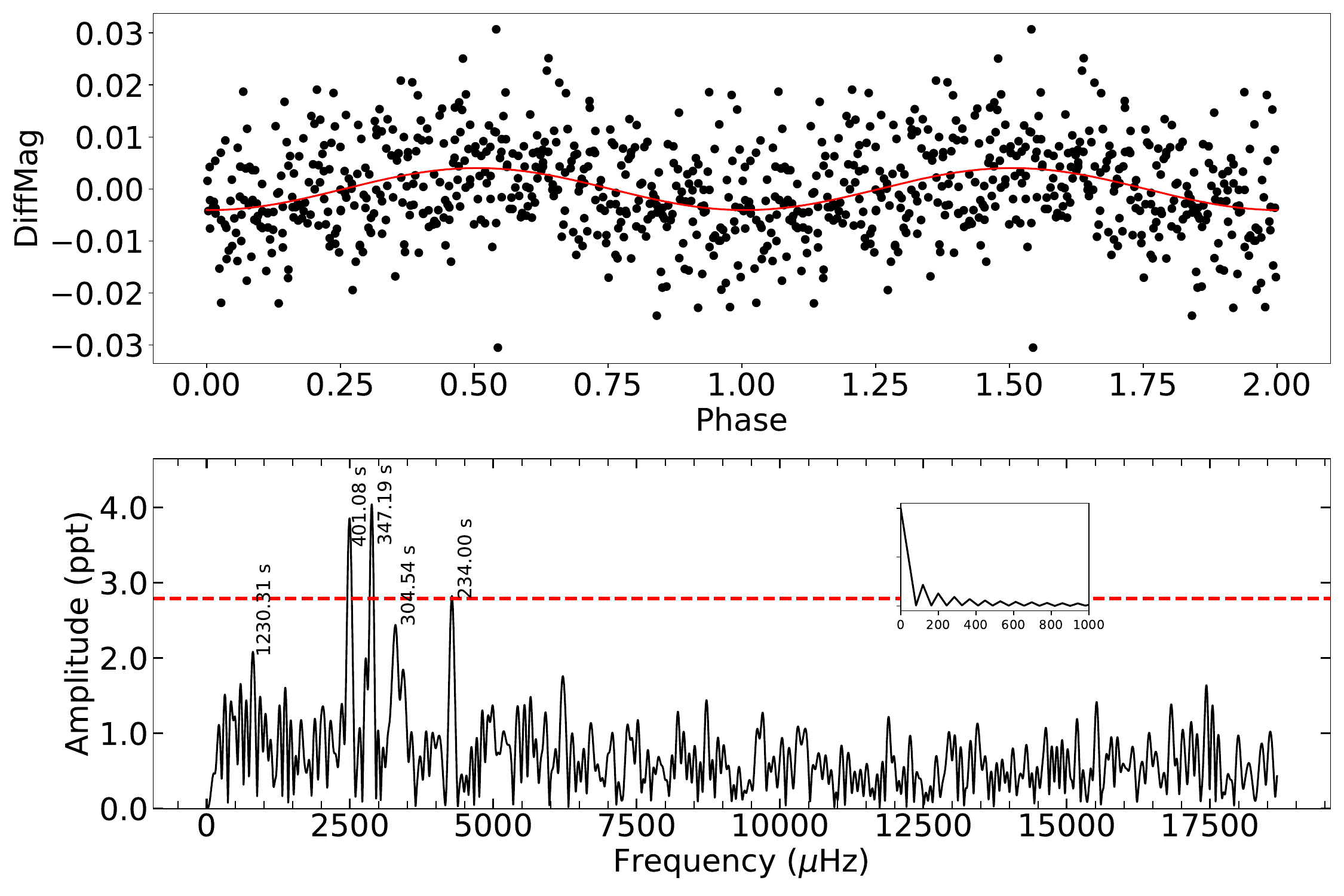}
  \caption{Top panel: Phased-folded light curve for the new pulsating low-mass WD SDSS J001245.60+143956.4. The light curve was folded to the highest amplitude pulsational period of about P=347\,s. The red solid line indicates a sinusoidal fit, depicted for better visualisation of the variation. Bottom panel: FT for the same star. The horizontal red dashed line indicates our detection limit, and the spectral window for each case is depicted as an inset plot. }
    \label{Fig:low-mass-FT_LC}
\end{center}
\end{figure}

In Figure \ref{Fig:low-mass-FT_LC} (top panel) we show the folded light curve with the period of P=347\,s for the SOAR night as well as its FT (bottom panel). Our analysis has shown five peaks above our threshold, being the highest amplitude peak corresponding to a period of about 347 seconds. The pulsation periods observed for this object are also consistent with the typical pulsation periods of low-mass WDs. For comparison, a pulsating low-mass WD identified by \citet{2020Parsons} within a binary system, SDSS J115219.99-024814.4, exhibits pulsation periods ranging from 1314\,s to 582\,s. Although the WD observed by \citet{2020Parsons} has a lower measured mass (0.325 $\pm$ 0.013 $M_{\odot}$) compared to the object analysed in this work (0.482 $\pm$ 0.03 ${M}_{\odot}$), the observed pulsation periods fall within the expected range for low-mass WDs \citep{2016Corscio_althaus, 2018Calcaferro}.

Furthermore, the location of this objects within $T_{\mathrm{eff}}$ --$\log g$ plane (see Figure \ref{Fig:TeffLogg_with_stars_marked}) also places this objects within the instability strip.

\begin{table}[h!]
\caption{Detected frequencies for the new pulsating low-mass WD SDSS J001245.60+143956.4.}             
\label{table:low-mass_new_freq_period_amp}      
\centering          
\begin{tabular}{c c c c }     
 Freq ($\mu Hz$)&	Amp (ppt)	&	P (sec)	&	FAP (ppt)	\\ 
 \hline\hline
 2880.278$\pm$5& 4.04$\pm$0.8 & 347.2$\pm$0.6 &2.324 \\
  2493.253$\pm$5&4.01$\pm$1 &  401.1$\pm$0.8& 2.376 \\ 
  4273.566$\pm$8&2.8$\pm$1  & 234.0$\pm$0.4 &  2.306\\
 3283.598$\pm$10& 2.5$\pm$0.4& 304.5$\pm$0.9 & 2.1718 \\ 
 810.714$\pm$13& 2.1 $\pm$0.5 & 1230.3$\pm$20 & 2.098 \\

\end{tabular}

{\raggedright \textbf{Notes:} The table lists the frequency, amplitude, period, and FAP detection limit for each peak. Observations were performed with SOAR. \par}
\end{table}

\subsubsection{SDSS J090559.60+084324.9} 

\begin{figure}[ht!]
\begin{center}
  \includegraphics[width=8cm, angle=0]{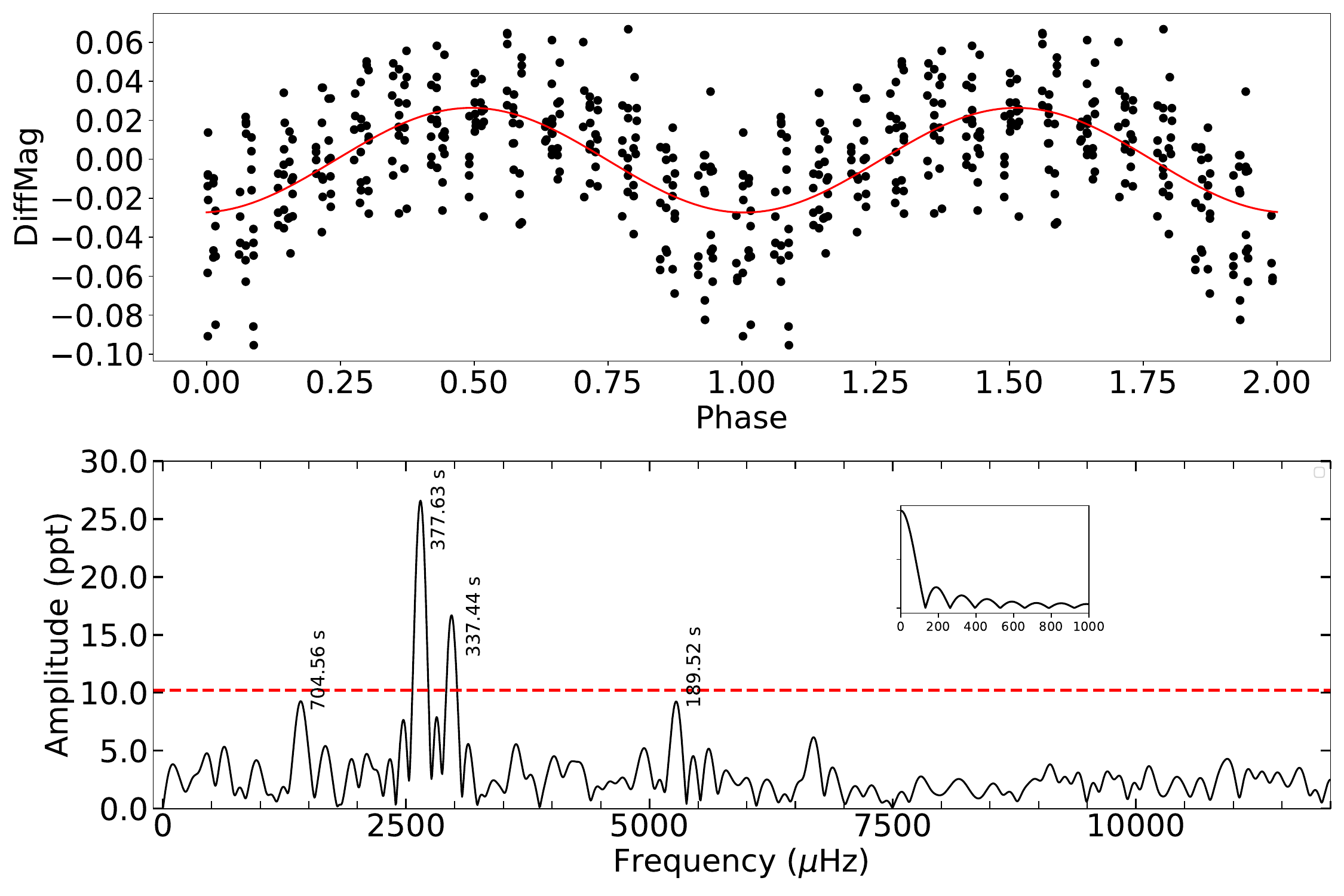}
  \caption{Top panel: Phased-folded light curve for the new ZZ Ceti SDSS J090559.60+084324.9. The light curve was folded to the highest amplitude pulsational period of about P=377\,s. The red solid line indicates a sinusoidal fit, depicted for better visualisation of the variation. The "lined" appearance in the folded light curve is attributed to the 337\,s pulsation period being close to an alias of the cadence. Bottom panel: FT for the same star. The horizontal red dashed line indicates our detection limit, and the spectral window is depicted as an inset plot.}
    \label{Fig:ZZCeti_LC_FT}
  
\end{center}
\end{figure}

SDSS J090559.60+084324.9 is the third pulsational object discovered in this work. It has a magnitude of $G=18.00$ and is located at a Gaia-derived distance of 1/parallax=$245 \pm 8$ pc. This star has been observed during 2.11\,hs with SOAR telescope, in which four periodical peaks were observed above the threshold (see bottom panel of Figure \ref{Fig:ZZCeti_LC_FT}). This star has also an SDSS spectrum available and our spectral fit has retrieved atmospheric parameters of $T_{\mathrm{eff}}$=10100$^{+630}_{-560}$K and $\log g$=8.2$\pm 0.4$. Once more, our spectral fit is in perfect agreement with the work done by \citet{2013Kleinman}, in which the authors have measured $T_{\mathrm{eff}}$=10070.0 $\pm$ 38, $\log g$=8.18 $\pm$ 0.04 and a mass of 0.704 $\pm$ 0.03	${M}_{\odot}$ (see Figure \ref{Fig:logg_teff-pulsation_comp_append}). Those values identified this star as a typical ZZ Ceti. 

For this object in particular, it was not surprising that we identified a ZZ Ceti star. In our candidate selection process, we accounted for the uncertainties in $T_{\mathrm{eff}}$ and $\log g$, which were considerable in many cases. Additionally, we extended our search to include slightly higher $T_{\mathrm{eff}}$ and $\log g$ values. This approach was motivated by the fact that the instability strips of ZZ Ceti and ELMV stars are a continuation of each other, with no clearly defined boundaries separating the two groups to date.

Nevertheless, the atmospheric parameters derived in this work place this new ZZ Ceti slightly outside the red edge of the instability strip. (see Figure \ref{Fig:TeffLogg_with_stars_marked}), towards the red edge. However, the star's location is still close to the positions of other known ZZ Ceti stars.

\begin{table}[h!]
\caption{Detected frequencies for the new ZZ Ceti SDSS J090559.60+084324.9.}             
\label{table:ZZCeti_new_freq_period_amp}      
\centering          
\begin{tabular}{c c c c }     
 Freq ($\mu Hz$)&	Amp (ppt)	&	P (sec)	&	FAP (ppt)	\\ 
 \hline\hline
2648.09348 $\pm$6  &   26.59 $\pm$9&   377.6$\pm$0.9   &6.831\\ 
2963.49915 $\pm$34  & 13.38 $\pm$9 & 337.4$\pm$4  &7.716\\
5276.4741  $\pm$14  & 9.04$\pm$1  &189.5$\pm$0.5& 7.618\\
1419.32554 $\pm$15  & 8.76$\pm$2  &   704.5$\pm$7 &6.885 \\

\end{tabular}

{\raggedright \textbf{Notes:} The table lists the frequency, amplitude, period, and FAP detection limit for each peak. Observations were performed with SOAR. \par}
\end{table}

\subsection{New binary white dwarfs}
\label{Section_new_binary}

\subsubsection{SDSS J045116.83+010426.6} 
\label{Section:J0451_strong_ellipsoidal}

\begin{figure}[t]
\begin{center}
  \includegraphics[width=8cm, angle=0]{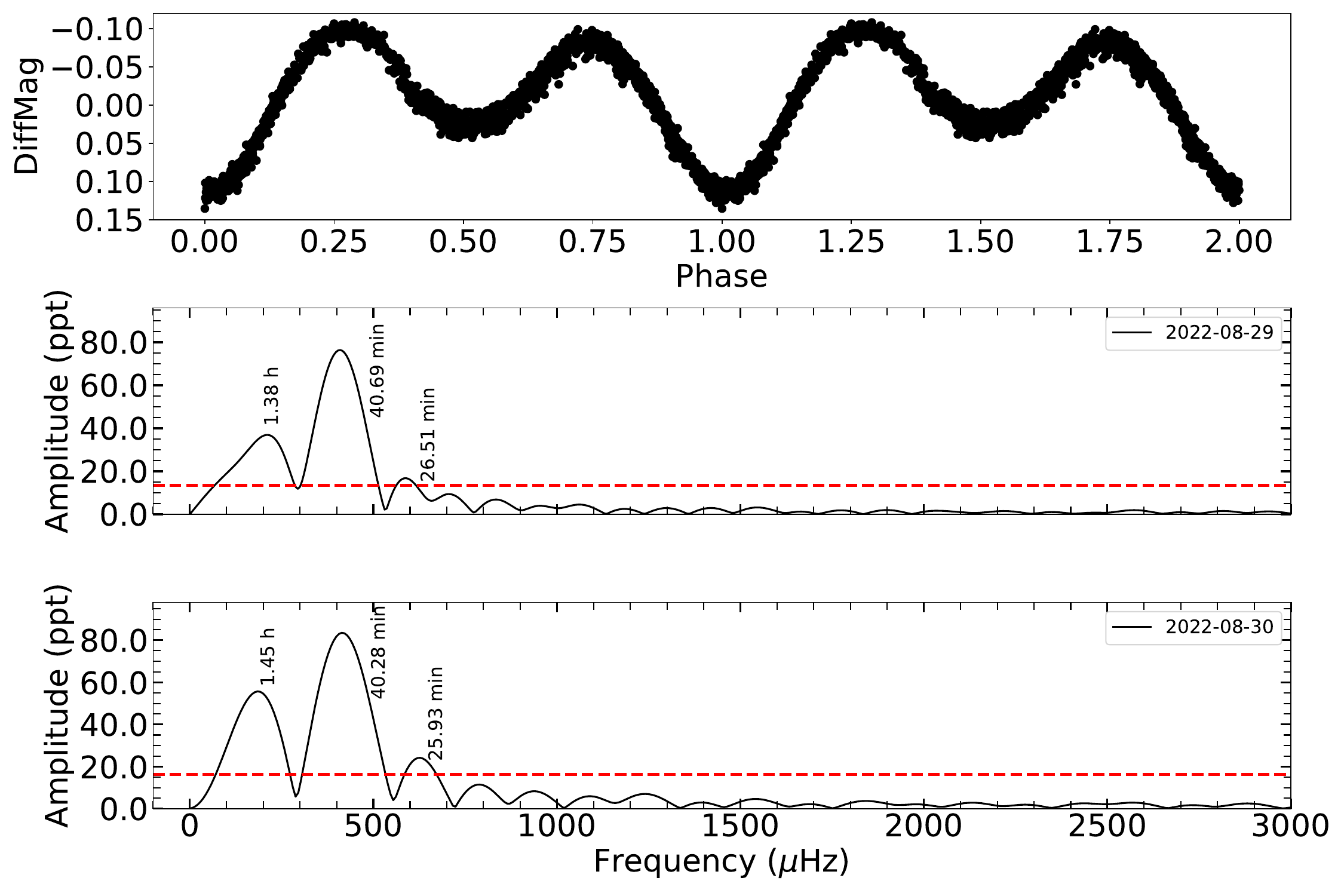}
  \caption{Top panel: SDSS J045116.83+010426.6 phased-folded light curve to $P_{orb}$=1.35\,h, which is twice the period of the peak with the highest amplitude in the FT. Middle and bottom panels show the FT for two different SOAR nights. The horizontal red dashed line in both FT indicates our detection limit. }
    \label{Fig:Binary_preELM}
  
\end{center}
\end{figure}

SDSS J045116.83+010426.6 has a magnitude of $G=15.35$ and is located at a Gaia-derived distance of 1/parallax=$294 \pm 3$ pc. This star has been observed in two different nights with SOAR, in which both light curves showed strong variations as seen in the phased-folded light curve in the top panel of Figure \ref{Fig:Binary_preELM} for just one of SOARs night. This variation in the light curve appears to show an ellipsoidal variation, presumably due to the gravitational deformation of one of the two stars \citep{2022Barlow}. Also, the significant difference between the minima in the light curve is striking and suggests that additional data, particularly in other filters, could provide insight into its origin. 

We have estimated the orbital period to be $P_{orb}$=1.35\,h for the combined nights, which is two times the highest peak in both FT (middle and bottom panel) in Figure \ref{Fig:Binary_preELM}. The $\sim$ 26\,min peak that appears in both periodograms of Figure \ref{Fig:Binary_preELM}, is a combination of the $\sim$ 1.4\,h and $\sim$ 40\,min peaks.

For this object we carried spectroscopic observations with the INT, which revealed a rich spectrum with strong metal lines. Furthermore, a preliminary SED and the co-added spectrum which reveals a significant presence of metals (as shown in Figure \ref{Fig:fit_chi}), suggest a primary star consistent with a pre-ELM, with $T_{\mathrm{eff}}$ = 7000$_{+200}^{-160}$ K and $\log g$=4.9$\pm$ 0.5.

A spectrum with higher signal-to-noise ratio (SNR) and higher resolution will be necessary to confirm the system orbital period with radial velocity (RV) measurements, to fully confirm the primary star nature, and to determine the nature of the secondary companion.


\subsubsection{SDSS J183245.52+141311.2} 

\begin{figure}[t]
\begin{center}
  \includegraphics[width=8cm, angle=0]{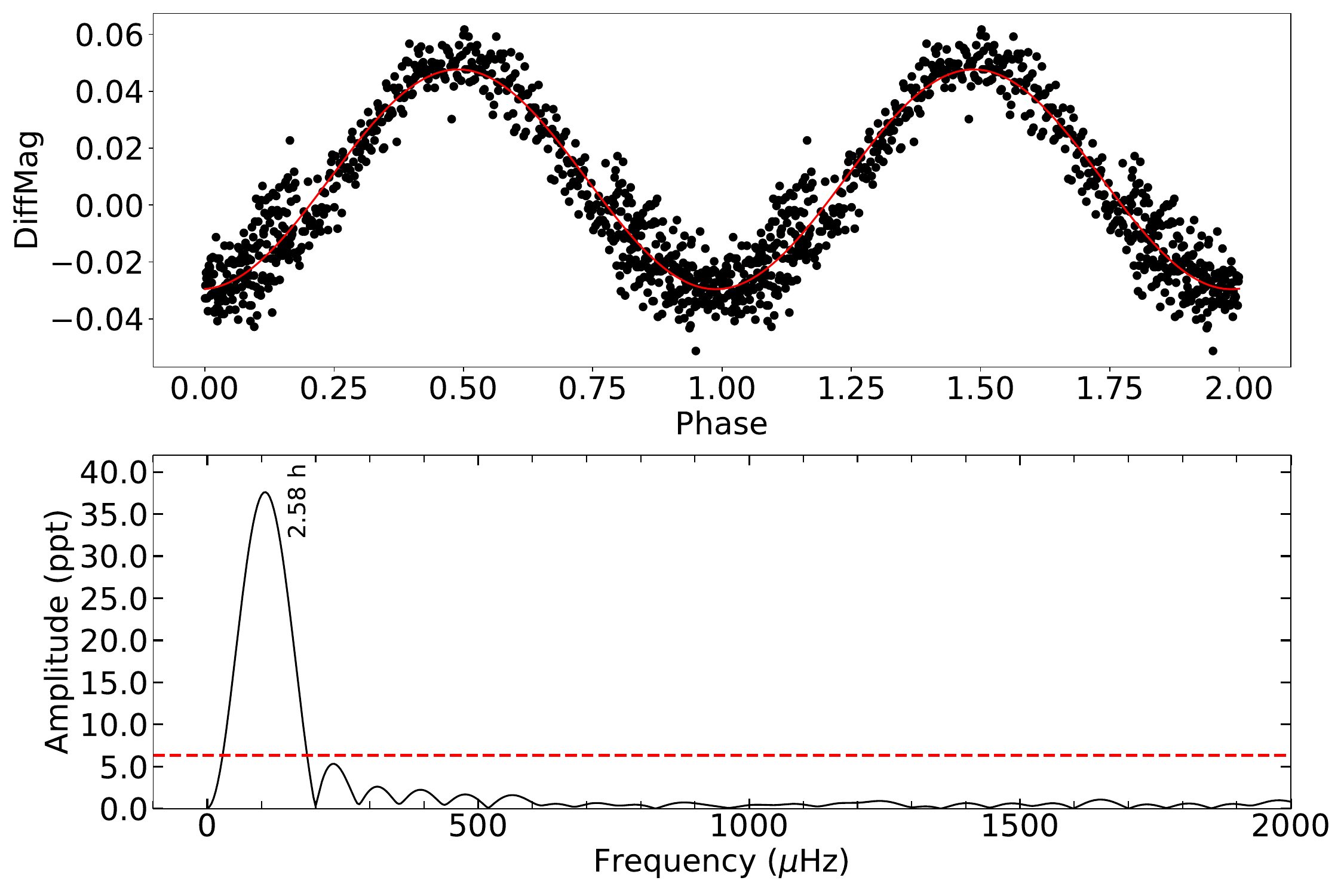}
  \caption{Top panel: SDSS J183245.52+141311.2 phased-folded light curve to $P_{orb}$=2.58\,h showing a strong reflection effect. The red solid line indicates a sinusoidal fit, depicted for better visualisation of the variation. Bottom panel shows the FT with a high amplitude peak high above our detection limit, with the latest being indicated by the red dashed line.}
    \label{Fig:J1832_reflection_FT_LC}
 
\end{center}
\end{figure}

SDSS J183245.52+141311.2 has been observed during only one night with SOAR, and it has a magnitude of $G=17.45$ and 1/parallax=$1503 \pm 207$ pc. The light curve of this object has shown a strong sinusoidal variation with an amplitude of about $\sim$ 0.05 mag (see top panel of Figure \ref{Fig:J1832_reflection_FT_LC}). The FT (bottom panel of Figure \ref{Fig:J1832_reflection_FT_LC}) depict a strong peak of P=2.58\,h. The high amplitude variation in its light curve, and its detected period is very characteristic for reflection effect, in which it originates from the irradiation of a cool companion by the hot primary star. The projected area of the companion’s heated hemisphere changes while it orbits the primary.
Our spectral fit indicates that the primary star has ($T_{\mathrm{eff}}$ = 8860$\pm 100$ K and $\log g$=5.9$\pm0.1$. The spectroscopic fit indicates the presence of additional light, suggesting a cooler companion, as evidenced by the slope of the spectrum shown in Figure \ref{Fig:fit_chi}. This interpretation aligns with the observed reflection effect in the system. However, the wavelength range of our spectra does not extend far enough into the red to reveal distinct features of a potential M dwarf. Consequently, we cannot rule out the possibility of a cool WD, a brown dwarf, or an M dwarf  as a companion.

Further spectroscopic follow-up is needed in order to obtain RV measurements, necessary in order to confirm the binary nature of this object, and to detected the nature of the secondary star.


\subsubsection{SDSS J083417.21-652423.2} 

\begin{figure}[t]
\begin{center}
  \includegraphics[width=8cm, angle=0]{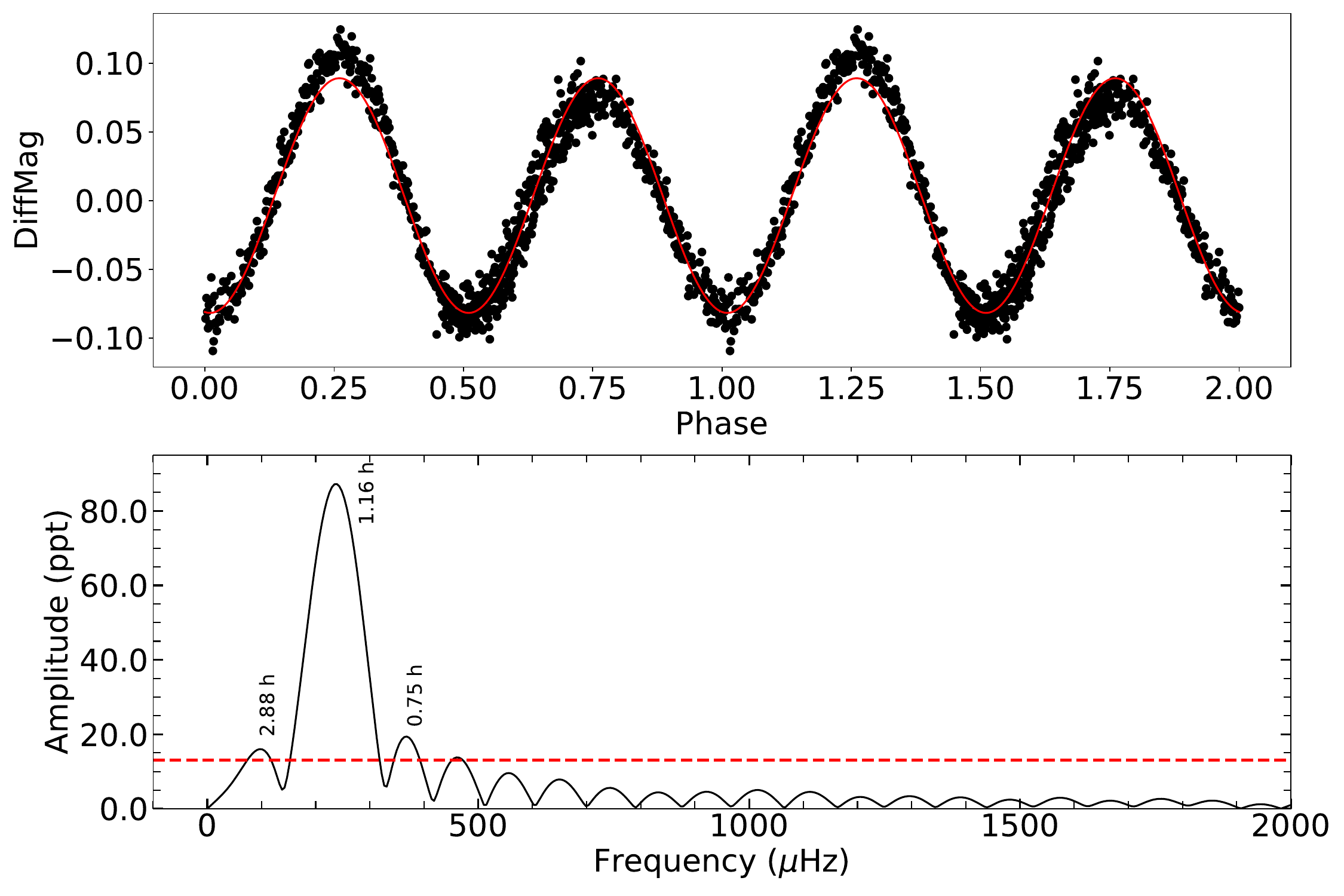}
  \caption{Top panel: SDSS J083417.21-652423.2 phased-folded light curve, showing the ellipsoidal variation, to $P_{orb}$=2.32\,h which is twice the period of the peak with the highest amplitude in the FT. The red solid line indicates a sinusoidal fit, depicted for better visualisation of the variation. Bottom panel shows the FT. The horizontal red dashed line indicates our detection limit.}
    \label{Fig:J0834_FT_LC}
 
\end{center}
\end{figure}

SDSS J083417.21-652423.2, $G=16.77$ and 1/parallax=$790 \pm 27$ pc, has been observed with SOAR during just one night, in which we have estimated the orbital period to be P=2.32\,h. The object light curve, shown in top panel in Figure \ref{Fig:J0834_FT_LC}, exhibit a characteristic, quasi-sinusoidal shape in which the flux peaks are slightly sharper than the valleys.


Similar to the case of the object SDSS J045116.83+010426.6 (see Section \ref{Section:J0451_strong_ellipsoidal}), the highest amplitude peak in the FT (bottom panel in Figure \ref{Fig:J0834_FT_LC}) represents half of the orbital period, since one can clearly see the difference in two consecutive peaks and valleys in the folded light curve (top panel in Figure \ref{Fig:J0834_FT_LC}) once we fold the light curve to twice the period of the main highest peak. 
The consecutive troughs and crests are uneven, which is likely due to a combination of effects, such as ellipsoidal modulation, gravity darkening and reflection reflect \citep{2022Barlow}.


Our spectral fit yielded $T_{\mathrm{eff}} = 7000^{+200}_{-160}$ K and $\log g = 4.9\pm 0.5$. However, further observations with higher SNR and resolution are needed to better constrain the atmospheric parameters and radial velocity measurements, and to confirm the orbital period of the system.

\subsection{Non observed to vary (NOV)}
\label{Section_nov}

From the observed sample, we did not detect any variability on the
FT for ten objects, within our detection limit (see table \ref{table:FAP_NOV_append}),
and thus they are classified as Not Observed to Vary (NOV). We
list the objects in Table \ref{table:variabilityInfo}, and their light curves and periodogram can be seen in Figures \ref{Fig:NOV_LC_mosaic_append} and \ref{Fig:NOV_FT_mosaic_append}, respectively. Since the low-mass and ELM instability strip is possibly not pure, we might have indeed objects that fall within the instability strip and have no pulsational variability, as well as we might have estimate atmospheric parameters wrongly. Furthermore, we recommend a follow-up observations for those objects given that  low-amplitude for pulsations for ELMs, which makes quite challenging to detect pulsations in those objects.



\section{Summary and conclusions}
In our comprehensive analysis, we observed 16 objects in high-speed photometry across multiple nights using the SOAR/Goodman and SMARTS-1m telescopes. This led to the detection of three new pulsating WDs: a new pulsating Extremely Low-Mass White Dwarf (ELMV), a low-mass white dwarf, and a ZZ Ceti star. 
 \begin{itemize}
     \item SDSS J212935.23+001332.3 is identified as a potential new pulsating ELM WD, with prominent pulsation periods of 1.22 hours and 42 minutes. Our spectral fitting of the observed INT spectra yields an $T_{\mathrm{eff}}$ of 9700$^{+480}_{-350}$ K , $\log g$ = 6.8$\pm$0.2, in which positioned this object close to known ELMVs in the $T_{\mathrm{eff}}$--$\log g$ plane. Along with a measured mass of 0.244 $\pm$ 0.002 \(\mathrm{M}_{\odot}\) \citep{2023Caron}, state this object as as new pulsating ELM. Further detailed spectroscopic analysis and asteroseismology analysis would be necessary in order to further confirm its ELMV nature, since the spectroscopical fit suggests additional light.  
    \item SDSS J001245.60+143956.4 exhibits pulsation with a dominant period of around 347 seconds. Combined with our spectral fitting results, $T_{\mathrm{eff}}$=10200$^{+700}_{-650}$K and $\log g$=7.5$\pm0.4$, and measured mass of 0.482 $\pm$ 0.03 ${M}_{\odot}$ \citep{2013Kleinman}, places this star within the instability strip for ZZ Ceti and ELMV WDs and classify this star as a new pulsating low-mass WD. 
    \item SDSS J090559.60+084324.9 is a new confirmed ZZ Ceti star, with four observed pulsational periods. The star lies near the red edge of the ZZ Ceti instability strip, with measured $T_{\mathrm{eff}}$ and log\text{g} of $T_{\mathrm{eff}}$=10100$^{+630}_{-560}$K and $\log g$=8.2$\pm0.4$. Along with a measured mass of 0.704 $\pm$ 0.03${M}_{\odot}$ \citep{2013Kleinman}, shows that this start has consistent characteristics of a typical ZZ Ceti.
 \end{itemize}


In addition, three new binary systems were discovered, two of which exhibit ellipsoidal variations due to gravitational deformation, while the third shows reflection effects from a heated companion. Spectral analyses provided insights into the primary stars in these systems, but further follow-up, including radial velocity measurements, is needed to confirm the nature of the companions.

 \begin{itemize}
    \item SDSS J045116.83+010426.6 shows in its light curves strong ellipsoidal variations presumably due to the gravitational deformation of one of the two stars. The estimated orbital period is 1.35 hours. A preliminary SED and co-added spectrum fit suggested a primary star consistent with a pre-ELM ($T_{\mathrm{eff}}$ = $7000_{-160}^{+200}$ K, and and $\log g = 4.9\pm 0.5$)
    \item SDSS J183245.52+141311.2 light curve exhibits a strong sinusoidal variation, likely due to a reflection effect from a cool companion being irradiated by a hot primary. The estimated orbital period is 2.58 hours, and the derived atmospheric parameters of $T_{\mathrm{eff}}$ = 8860$\pm 100$ K and $\log g$=5.9$\pm0.1$ suggests the primary star to be an WD. However, further spectroscopic follow-up is needed to confirm the binary nature and the secondary star properties.
    \item SDSS J083417.21-652423.2 light curve suggests ellipsoidal variations, and the uneven peaks and valleys indicate gravity-darkening and Doppler beaming effects. Our analysis return an potential orbital period of 2.32 hours and $T_{\mathrm{eff}} = 7000^{+200}_{-160}$ K and $\log g = 4.9\pm 0.5$. Higher-quality data is needed to refine the orbital parameters and confirm the system’s details.
 \end{itemize}

Finally, ten other stars in the sample showed no detectable variability. While these objects fall within potential instability strips, they did not exhibit pulsations during the observation period. The absence of variability in these cases could be due to low-amplitude pulsations or inaccuracies in the atmospheric parameter estimates. Follow-up observations are recommended, as the detection of low-amplitude pulsations may have been hindered.



\begin{acknowledgements}
LAA acknowledges financial support from CONICYT Doctorado Nacional in the form of grant number No: 21201762 and ESO studentship program. IP acknowledges support from The Royal Society through a University Research Fellowship (URF/R1/231496).
Based on observations collected with the GMOS spectrograph on the 8.1~m Gemini-South telescope at Cerro Pachón, Chile, under the program GS-2022A-Q-416, and observation with Goodman SOAR 4.1~m telescope, at Cerro Pachón, Chile, and SMARTS-1m telescope at Cerro Tololo, Chile, under the program allocated by the Chilean Time Allocation Committee (CNTAC), no: CN2022A-412490, and SOAR observational time through NOAO programs 2021A-XXXX. 
M.V. acknowledges support from FONDECYT (grant No: 1211941).
\end{acknowledgements}

%
%

\bibliographystyle{aa} 
\bibliography{bibi} 


\begin{appendix}
\onecolumn

\section{Details of the photometric and spectroscopic observations}

\renewcommand{\arraystretch}{1.5}
\begin{table*}[h!]
\caption{Journal of photometric observations from ground-based facilities.}             
\label{table:phot_log}      
\centering          
\begin{tabular}{c c c c c c}     
SDSS J	&	Start (BJD-2400000)	&	Duration(hs)	&	Cadence(sec)	&	Exposure time (sec)	&	Telescope	 \\ 
 \hline\hline

001245.60+143956.4	&	59853.61862	&	3.4	&	26	&	20	&	SOAR	\\ \hline 
\multirow{2}{*}{002602.29-103751.9}&	59820.7012	&	3.74	&	32	&	28	&	SMARTS 	\\  
&	59854.529 &	2.04	&	11	&	5	&	SOAR	\\ \hline 
015958.50-180546.8	&	59854.67075	&	2.11	&	20	&	15	&	SOAR	\\ \hline 

040600.99-542751.7	&	59899.62891	&	2.4	&	17	&	10	&	SOAR	 \\ \hline 

\multirow{2}{*}{045116.83+010426.6}&	59821.80796	&	2.4	&	12	&	5	&	SOAR	\\
&	59822.8369	&	1.89	&	12	&	5	&	SOAR	 \\ \hline 

083013.72-100356.0	&	59645.51678	&	2.3	&	32	&	25	&	SOAR 	\\ \hline 

083417.21-652423.2	&	59898.7132	&	3	&	12	&	6	&	SOAR	 \\ \hline 
090559.60+084324.9 	&	59997.59597	&	2.11	&	26	&	20	&	SOAR	 \\ \hline
090901.01-454637.0	&	59649.49265	&	1	&	22	&	15	&	SOAR	\\ \hline

\multirow{2}{*}{112521.26-023433.4}&	59644.6176	&	2.32	&	67	&	60	&	SOAR	\\
&	59649.63528	&	3.43	&	67	&	60	&	SOAR	 \\ \hline 
\multirow{2}{*}{134619.11-135026.7}&	59691.61378	&	0.43	&	26	&	22	&	SMARTS	\\
&	59725.48814	&	4.29	&	45	&	40	&	SMARTS	\\ \hline 
\multirow{2}{*}{151626.39-265836.9}&	59703.68379	&	2.24	&	63	&	60	&	SMARTS	\\
&	59704.62611	&	4	&	65	&	60	&	SMARTS	\\ \hline

\multirow{2}{*}{161004.27+063330.6}&	59645.74481	&	3.45	&	97	&	80	&	SOAR	\\
&	59649.7782	&	2.37	&	67	&	60	&	SOAR	 \\ \hline 
183245.52+141311.2	&	59821.49728	&	3.61	&	18	&	12	&	SOAR	\\ \hline 

\multirow{2}{*}{183702.03-674141.1}&	59725.66987	&	1.9	&	64	&	60	&	SMARTS	\\
&	59854.47724	&	1	&	16	&	10	&	SOAR	\\ \hline 
\multirow{4}{*}{212935.23+001332.3}	&	59821.65449	&	2.22	&	12	&	5	&	SOAR	\\
&	59853.48298     &	1.63	&	12	&	5	&	SOAR	\\
&	59819.54055	&	3.76	&	33	&	28	&	SMARTS \\
&	59898.50753	&	1.85	&	11	&	4	&	SOAR	 \\\hline
 \hline
\end{tabular}

{\raggedright \textbf{Notes:} We list the SDSS J identifier, start of the run in BJD, duration of each run, cadence time, exposure time and telescope used in columns 1, 2, 3, 4, 5 and 6, respectively. \par}
\end{table*}

\begin{table*}[h!]
\caption{Details of the spectroscopic observations that contributed to this work.}             
\label{table:spec}      
\centering          
\begin{tabular}{c c c c c c}     
 \hline\hline
Telescope/spectrograph      &       Grating       &          Wavelength range      & Typical R $(\frac{\lambda}{\Delta \lambda})$  & Stars observed & Proposals \\ 
\hline
\multirow{5}{*}{SDSS} &  \multirow{5}{*}{-} & \multirow{5}{*}{3800-9200~\AA} & \multirow{5}{*}{1500-2500}  &  J001245.60+143956.4 & \multirow{5}{*}{-} \\
                      &                     &                  &                    & J002602.29-103751.9               &  \\
                      &                     &                  &                    & J090559.60+084324.9               &  \\
                      &                     &                  &                    & J112521.26-023433.4               &  \\ 
                      &                     &                  &                    & J161004.27+063330.6               &  \\
\hline                      
\multirow{6}{*}{SOAR/Goodman} & \multirow{6}{*}{930 l/mm}  &  \multirow{6}{*}{3600-5200~\AA}  & \multirow{6}{*}{$\sim 2000$}  &  J015958.50-180546.8  &  \\
                      &                     &                  &                    & J040600.99-542751.7              & SO2019A-005  \\
                      &                     &                  &                    & J083417.21-652423.2              & SO2019B-012 \\                      
                      &                     &                  &                    & J090901.01-454637.0              & SO2020B-010 \\
                      &                     &                  &                    & J151626.39-265836.9              & SO2021A-008 \\
                      &                     &                  &                    & J183702.03-674141.1              &  \\   
\hline
\multirow{4}{*}{INT/IDS} & \multirow{4}{*}{EEV10/1200B}  &  \multirow{4}{*}{3800-5000~\AA}  & \multirow{4}{*}{$\sim 2500$}  & J045116.83+010426.6  &  \multirow{4}{*}{i20an004} \\
                      &                     &                  &                    & J134619.11-135026.7               &  \\
                      &                     &                  &                    & J183245.52+141311.2               &  \\                      
                      &                     &                  &                    & J212935.23+001332.3               &  \\                                            
\hline
Gemini-S/GMOS      &      B600               &   3600-6800~\AA  &   $\sim 1000$  & J083013.72-100356.0      &  GS-2020A-Q-315 \\ 

 \hline\hline
\end{tabular}
\end{table*}

\begin{table*}[h!]
\caption{List of the 16 objects  presented in this work.}             
\label{table:variabilityInfo}      
\centering       
\begin{threeparttable}
\begin{tabular}{c c c c c c}     
\hline     
SDSS J	&	$T_{\mathrm{eff}}$ (K) &	$\log g$(cgs)	&	Mass ($M_{\odot}$)	&	Variability	&	Reference \tnote{a} \\\hline \hline

\multirow{2}{*}{001245.60+143956.4}	&	10893 $\pm$ 68	&	7.74 $\pm$ 0.07	&0.482 $\pm$ 0.03	&	 \multirow{2}{*}{Pulsations} &	\citet{2013Kleinman}\\
& $10200^{+700}_{-650}$	&	$7.5\pm0.4$	&	-	&		&	This work, $\chi^2_\mathrm{red}=1.6$\tnote{c}	\\\hline
\multirow{2}{*}{002602.29-103751.9	}&	9177 $\pm$ 30	&	7.26 $\pm$ 0.01	&0.273 $\pm$ 0.004 &	\multirow{2}{*}{NOV	}&	\citet{2023Caron}	\\
&	$10000^{+480}_{-400}$&	$7.9\pm0.4$&	-	&		&	This work, $\chi^2_\mathrm{red}=4.1$	\\\hline
015958.50-180546.8	&	$7500^{+270}_{-290}$	&	$6.4\pm0.4$	&	-	&	NOV	&		This work, $\chi^2_\mathrm{red}=127\tnote{b}$	\\ \hline
040600.99-542751.7	&	$9300^{+300}_{-270}$	&	$7.0\pm0.2$	&	-	&	NOV	&		This work, $\chi^2_\mathrm{red}=122\tnote{b}$	\\ \hline
045116.83+010426.6	&	$7000^{+200}_{-160}$	&	$4.9\pm0.5$	&	-	&	Ellipsoidal 	&	This work,  $\chi^2_\mathrm{red}=245\tnote{d}$	\\ \hline
083013.72-100356.0	&	$7750\pm190$	&	$5.2\pm0.4$	&	-	&	NOV	&		This work,  $\chi^2_\mathrm{red}=84\tnote{d}$	\\ \hline
083417.21-652423.2	&	$7700^{+200}_{-170}$	&	$5.4\pm0.3$	&	-	&	Ellipsoidal 	&		This work, $\chi^2_\mathrm{red}=354\tnote{d}$	\\ \hline
\multirow{2}{*}{090559.60+084324.9} 	&	10070.0 $\pm$ 38	&8.18 $\pm$ 0.04 &0.704 $\pm$ 0.03 &	\multirow{2}{*}{Pulsations} &	\citet{2013Kleinman}     \\
&$10100^{+630}_{-560}$&	$8.2\pm0.4$&	-	&		&	This work, $\chi^2_\mathrm{red}=1.9$\tnote{c}	\\\hline
090901.01-454637.0	&	$9700^{+430}_{-330}$	&	$7.3\pm0.2$	&	-	&	NOV	&		This work,  $\chi^2_\mathrm{red}=18$	\\ \hline

\multirow{2}{*}{112521.26-023433.4}	&	 9615 $\pm$149&	 8.08 $\pm$0.2&	 0.644 $\pm$0.1&	\multirow{2}{*}{NOV}	&	\citet{2013Kleinman} \\
&	$8650^{+940}_{-900}$	&	$8.5\pm0.5$	&	-	&		&	This work, $\chi^2_\mathrm{red}=1.5\tnote{c}$	\\\hline
\multirow{2}{*}{134619.11-135026.7	}&	9102 $\pm$ 39	&	7.02 $\pm$ 0.009	&	0.216 $\pm$ 0.003	&	\multirow{2}{*}{NOV}	&	\citet{2023Caron}	\\
&	$10100^{+210}_{-380}$	&	$7.1\pm0.2$		&	-	&		&	This work,  $\chi^2_\mathrm{red}=136\tnote{b}$	\\\hline

151626.39-265836.9	&	$10100^{+880}_{-760}$	&	$7.9\pm0.4$	&	-	&	NOV	&	This work, $\chi^2_\mathrm{red}=1.7\tnote{c}$	\\\hline 
\multirow{2}{*}{161004.27+063330.6}	&	 9454$\pm$85&	 8.02 $\pm$0.1&	 0.61 $\pm$0.07&	\multirow{2}{*}{NOV}	&	\citet{2013Kleinman}\\ 
&	$8400\pm840$&	$8.5\pm0.5$&	-	&		&	This work, $\chi^2_\mathrm{red}=2.2\tnote{c}$	\\\hline
183245.52+141311.2	&	$8860\pm100$	&	$5.9\pm0.1$	&	-	&	Reflection effect	&		This work, $\chi^2_\mathrm{red}=1359\tnote{b}$ 	\\ \hline

183702.03-674141.1	&	$7750^{+190}_{-170}$	&	$5.2\pm0.4$	&	-	&	NOV	&		This work,  $\chi^2_\mathrm{red}=148\tnote{d}$	\\ \hline

\multirow{2}{*}{212935.23+001332.3	}&8860 $\pm$ 30 &7.16 $\pm$ 0.006 &0.244 $\pm$ 0.002 &	\multirow{2}{*}{Pulsations }	&	\citet{2023Caron}	 	\\
&$9700^{+480}_{-350}$	&	$6.8\pm0.2$	&	-	&		&	This work, $\chi^2_\mathrm{red}=232\tnote{b}$	\\\hline
\hline
\end{tabular}
{\raggedright \textbf{Notes:} Column 1 provides the SDSS J identifier. The effective temperature, surface gravity, and stellar mass determinations are listed in columns 2, 3, and 4, respectively. Column 5 shows the type of variability seen in each object's light curve. Column 6 contains the reference for the atmospheric parameters, along with the reduced $\chi^2$ for the parameters obtained in this work. \par}
\begin{tablenotes}\footnotesize
\item [a] References for $T_{\mathrm{eff}}$ (K),	$\log g$(cgs)	and Mass ($M_{\odot}$) fit values.
\item [b] Possible extra light.
\item [c] MCMC fit.
\item [d] Metals in the atmosphere.
\end{tablenotes}
\end{threeparttable}
\end{table*}

\twocolumn

\section{NOV: Extra information}
\label{appendix:NOV_extra}
\begin{table}[h!]
\caption{ List of 10 objects identified as NOV.}             
\label{table:FAP_NOV_append}      
\centering          
\begin{tabular}{c c  }     
SDSS J&	FAP (ppt)	\\ 
 \hline\hline
002602.29-103751.9 & 4.4954\\ 
112521.26-023433.4 & 7.0017 \\ 
161004.27+063330.6 & 4.5990\\ 
040600.99-542751.7 & 6.8635\\ 
015958.50-180546.8 & 13.8596\\ 
090901.01-454637.0 & 6.1749\\ 
083013.72-100356.0 & 3.1355\\ 
183702.03-674141.1 & 2.9153\\ 
134619.11-135026.7 & 8.6157\\ 
151626.39-265836.9 & 10.8188\\ \hline
\end{tabular}

{\raggedright \textbf{Notes:} The table includes the FAP detection limit for each star. \par}
\end{table}

  

\begin{figure*}[htbp]
\centering
   \includegraphics[width=0.85\textwidth]{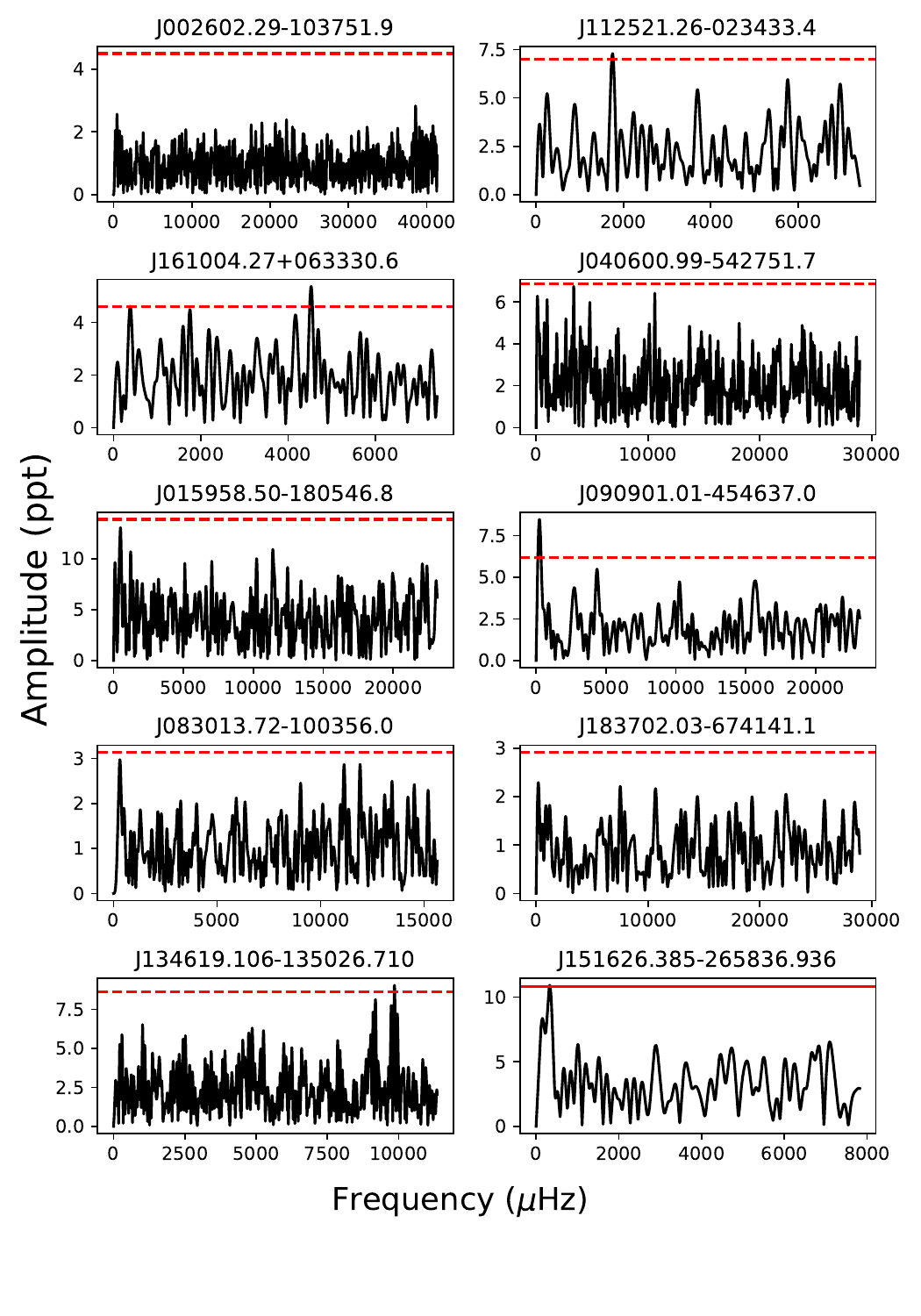}
     \caption{ FT of the 10 objects that were identified as NOV and their respective FAP detection limit depicted as red dashed line. }
     \label{Fig:NOV_FT_mosaic_append}
\end{figure*}

\clearpage

\begin{figure*}[htbp]
   \centering
   \includegraphics[width=0.85\textwidth]{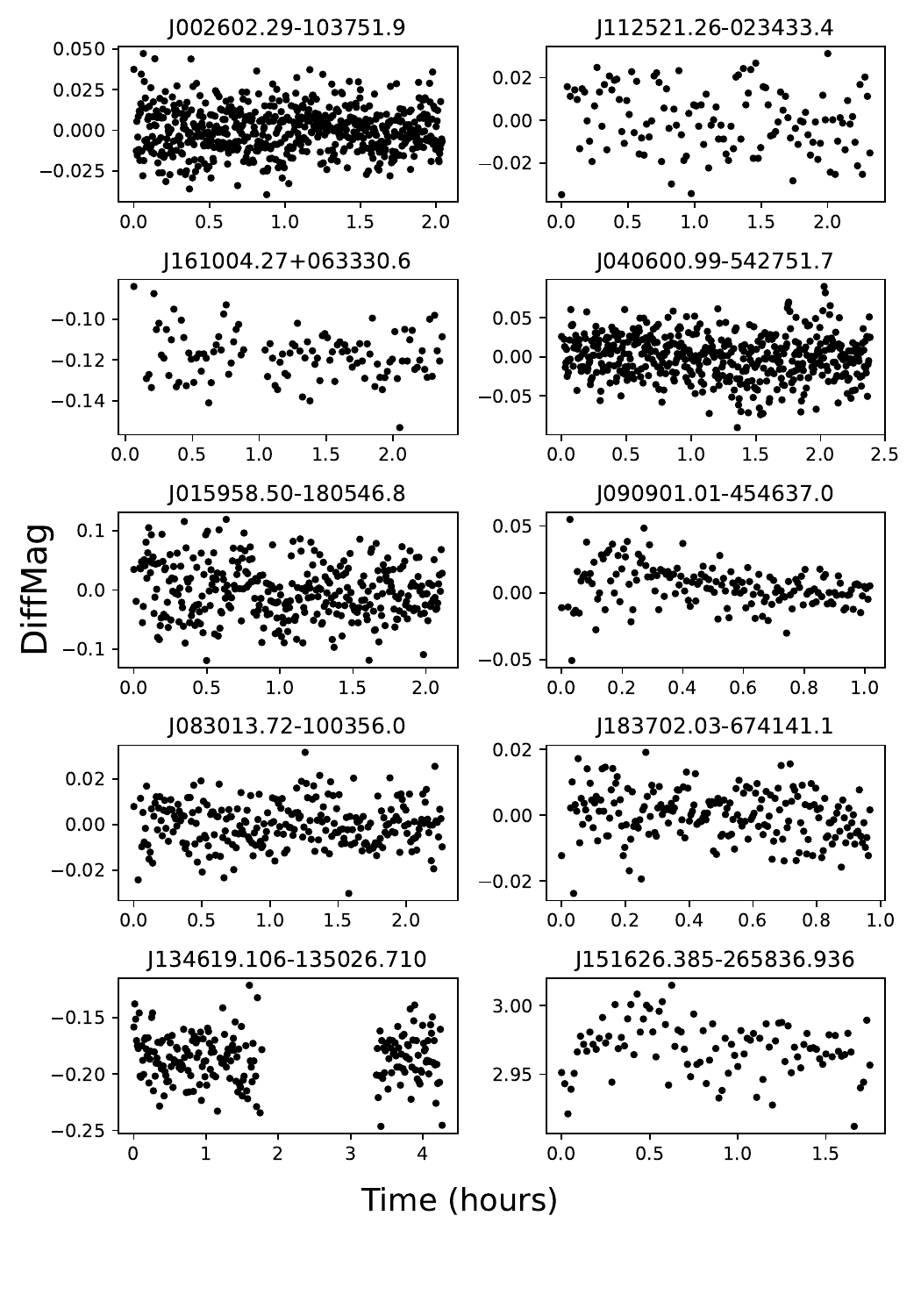}
      \caption{Light curve of the 10 objects that were identified as NOV.}
\label{Fig:NOV_LC_mosaic_append}
\end{figure*}

  

\clearpage

\section{Full spectral fits}
\label{appendix:spectra}

\begin{figure}[htbp]
\centering
   \includegraphics[width=0.5\textwidth]{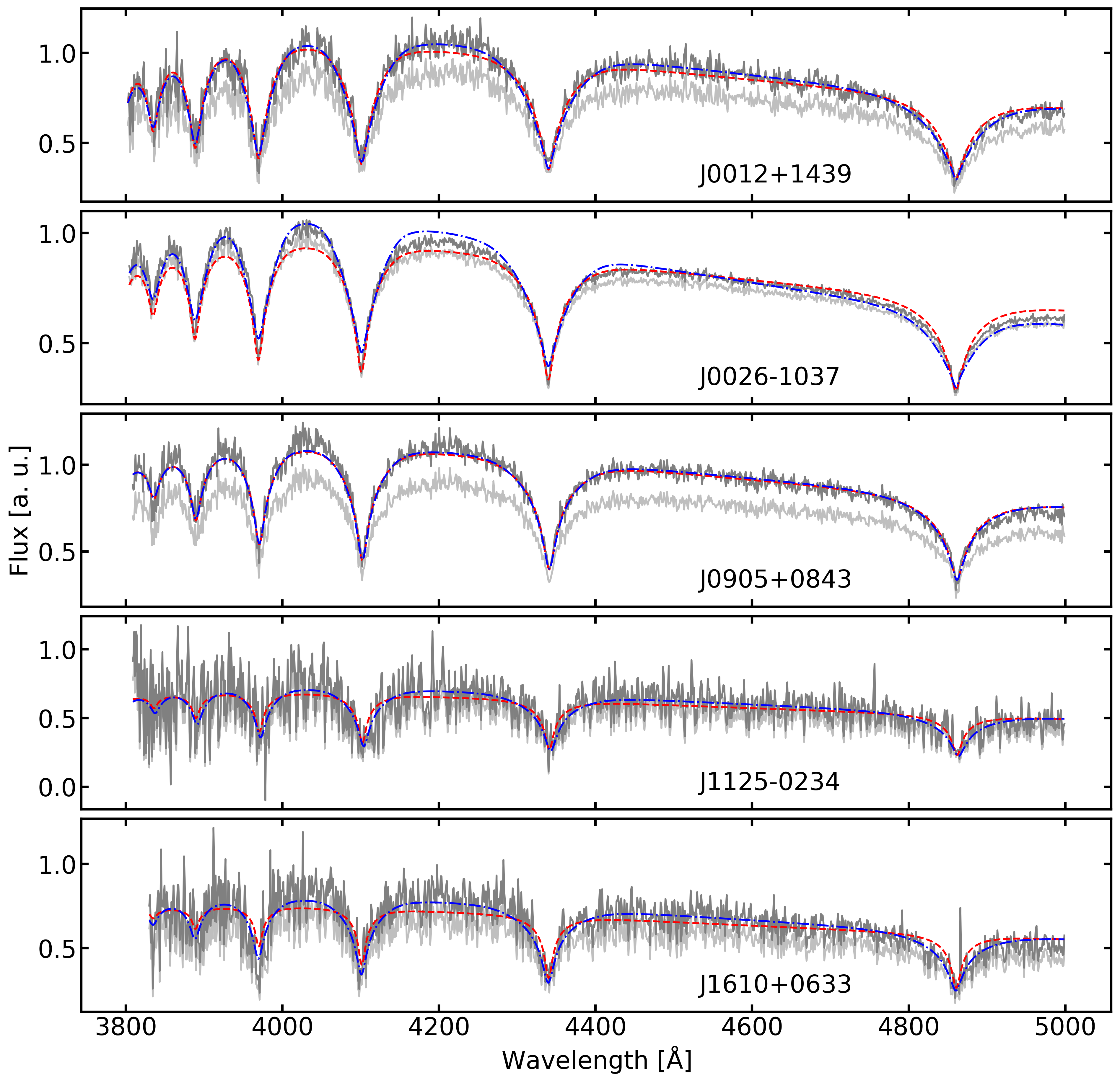}
     \caption{Spectral fits to objects that were fit using MCMC due to showing two minima in $\chi^2$. The observed spectrum is shown in light grey, the extinction-corrected spectrum in darker grey, the model for minimum $\chi^2$ in blue, and the adopted solution in red. }
     \label{Fig:fit_mcmc}
\end{figure}

\begin{figure}[htbp]
\centering
   \includegraphics[width=0.5\textwidth]{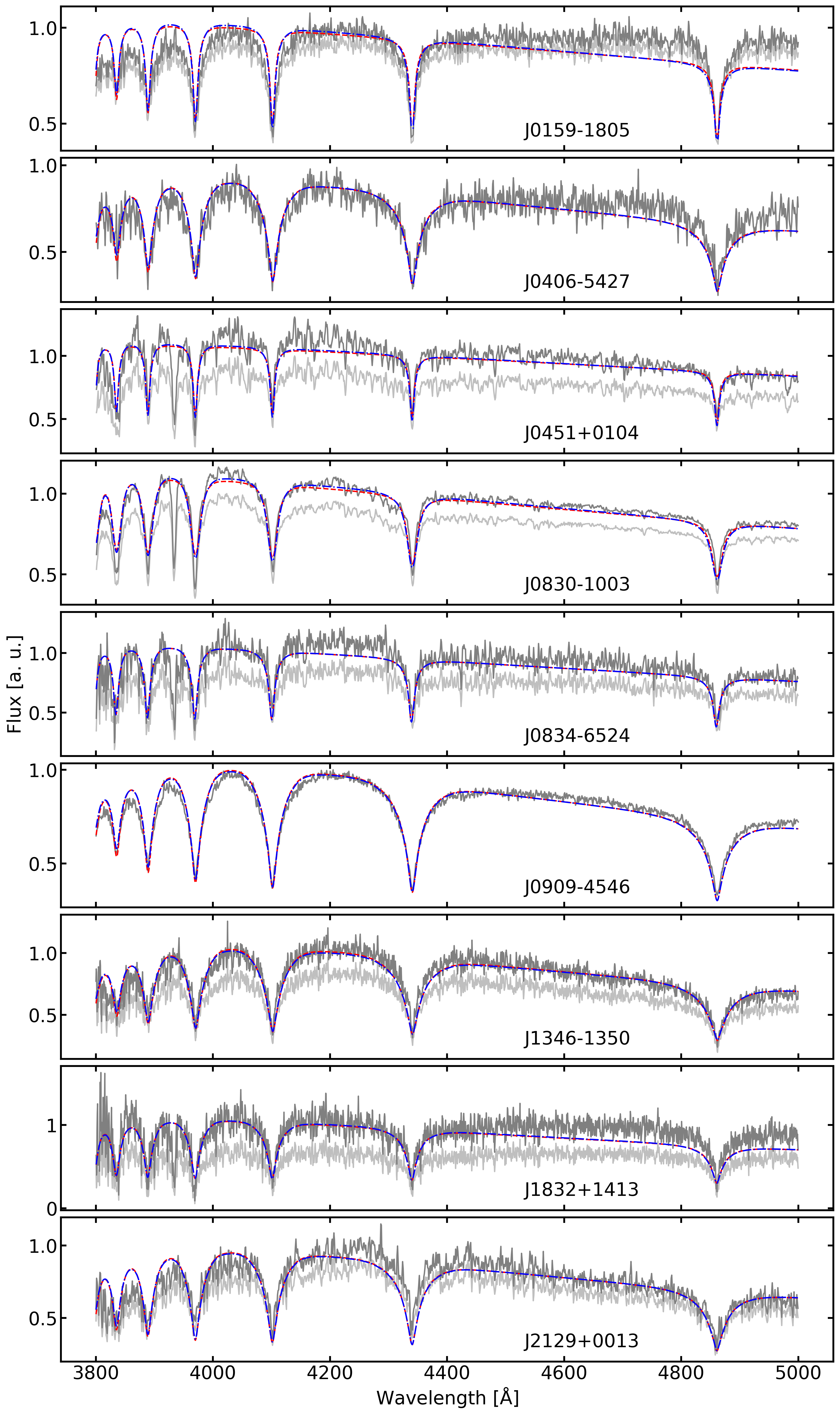}
     \caption{Spectral fits to objects with a single minimum in $\chi^2$. The observed spectrum is shown in light grey, the extinction-corrected spectrum in darker grey, the model for minimum $\chi^2$ in blue, and the adopted solution in red. }
     \label{Fig:fit_chi}
\end{figure}

\clearpage

\section{ Extra figures}
\label{appendix:extra}
\begin{figure}[ht!]
\begin{center}
  \includegraphics[width=0.5\textwidth]{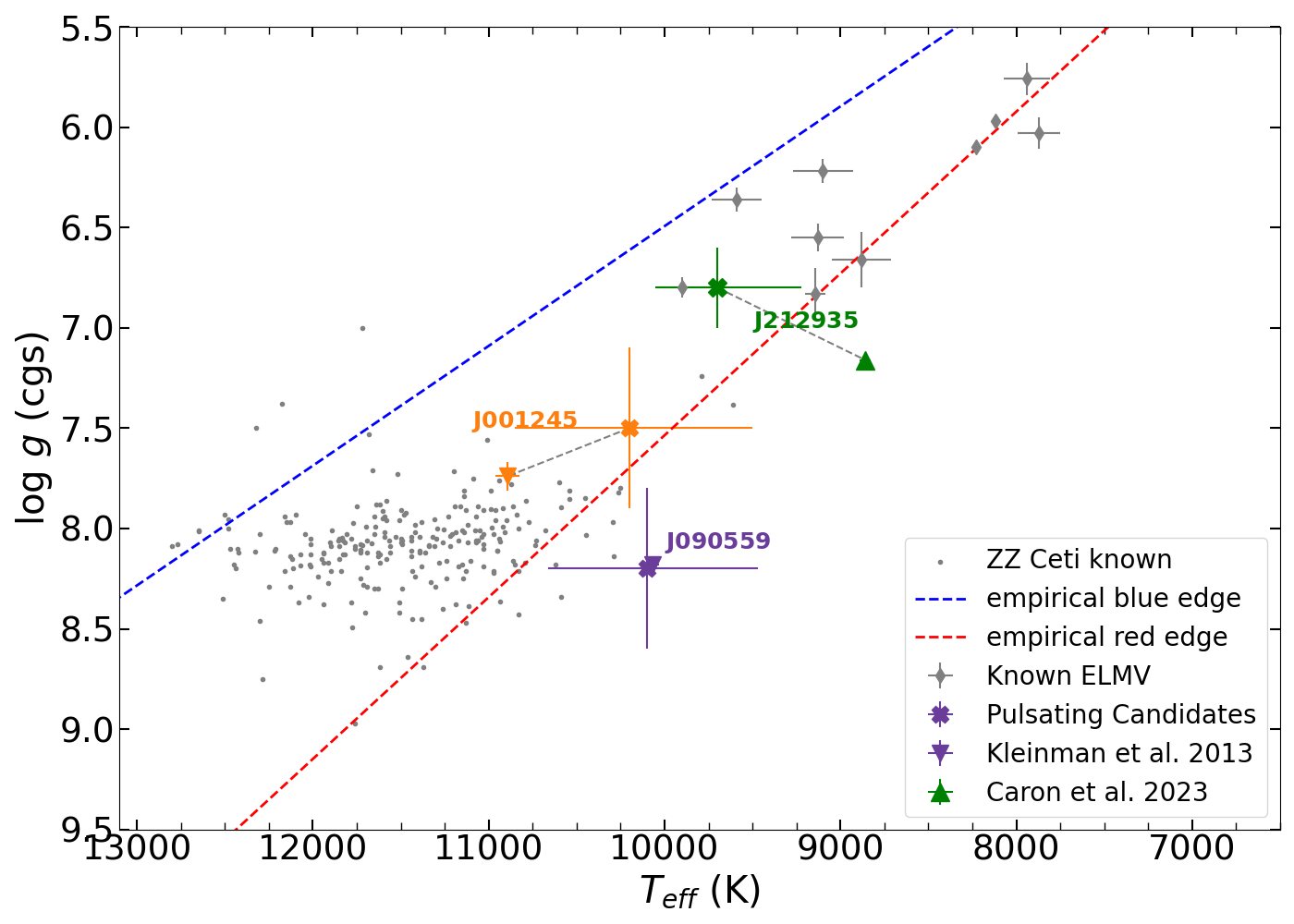}
  \caption{ $T_{\mathrm{eff}}-\log{g}$ plane shows the positions of the three new pulsating WDs identified in this work, with atmospheric parameters from this study marked by $\times$ symbols. For comparison, atmospheric parameters from \citet{2013Kleinman} and \citet{2023Caron} are indicated by upside-down triangle and triangle symbols, respectively. The known ZZ Ceti \citep{2022MRomero} and ELMVs \citep{2012Hermes, 2013Hermesa, 2013Hermesb, 2015Kilic, 2015Bell, 2017Bell, 2018Pelisolib, 2021Lopez} are shown as gray dots and gray diamond shapes, respectively. The empirical ZZ Ceti instability strip published in \citet{2015Gianninas_ELM6} is marked with dashed lines.}
    \label{Fig:logg_teff-pulsation_comp_append}
  
\end{center}
\end{figure}

\end{appendix}

\end{document}